\documentclass[sigconf]{acmart}

\usepackage{graphicx}
\usepackage{wrapfig}

\AtBeginDocument{%
  \providecommand\BibTeX{{%
    \normalfont B\kern-0.5em{\scshape i\kern-0.25em b}\kern-0.8em\TeX}}}

\copyrightyear{2023} 
\acmYear{2023} 
\setcopyright{acmlicensed}\acmConference[CHI '23]{Proceedings of the 2023 CHI Conference on Human Factors in Computing Systems}{April 23--28, 2023}{Hamburg, Germany}
\acmBooktitle{Proceedings of the 2023 CHI Conference on Human Factors in Computing Systems (CHI '23), April 23--28, 2023, Hamburg, Germany}
\acmPrice{15.00}
\acmDOI{10.1145/3544548.3580912}
\acmISBN{978-1-4503-9421-5/23/04}

\def\markup{1}

\if\markup1

\else

\newcommand{\st}[1]{}
\fi
\usepackage{soul}

\usepackage{hyperref}

\begin{document}

\title{StoryChat: Designing a Narrative-Based Viewer Participation Tool for Live Streaming Chatrooms}

\author{Ryan Yen}
\orcid{0001-8212-4100}

\affiliation{%
  \institution{City University of Hong Kong}
  \streetaddress{83 Tat Chee Ave}
  \city{Hong Kong}
  \country{China}
}
\affiliation{%
  \institution{University of Waterloo}
  \streetaddress{200 University Ave W}
  \city{Waterloo}
  \state{Ontario}
  \country{Canada}
}
\email{ryanyen2-c@my.cityu.edu.hk}

\author{Li Feng}
\orcid{0002-6198-0896}
\affiliation{%
  \institution{City University of Hong Kong}
  \streetaddress{83 Tat Chee Ave}
  \city{Hong Kong}
  \country{China}
}
\affiliation{%
  \institution{The Hong Kong University of Science and Technology (Guangzhou)}
  \streetaddress{83 Tat Chee Ave}
  \city{Guangzhou}
  \country{China}
}
\email{lfeng256@connect.hkust-gz.edu.cn}

\author{Brinda Mehra}
\orcid{0002-0606-8191}

\affiliation{%
  \institution{City University of Hong Kong}
  \streetaddress{83 Tat Chee Ave}
  \city{Hong Kong}
  \country{China}
}
\affiliation{%
  \institution{University of Michigan Ann Arbor}
  \city{Ann Arbor}
  \state{Michigan}
  \country{USA}
}
\email{brinda@umich.edu}

\author{Ching Christie Pang}
\orcid{0003-4704-2403}

\affiliation{%
  \institution{City University of Hong Kong}
  \streetaddress{83 Tat Chee Ave}
  \city{Hong Kong}
  \country{China}
}
\email{chinpang8-c@my.cityu.edu.hk}

\author{Siying Hu}
\orcid{0002-3824-2801}

\affiliation{%
  \institution{City University of Hong Kong}
  \streetaddress{83 Tat Chee Ave}
  \city{Hong Kong}
  \country{China}
}
\email{siyinghu-c@my.cityu.edu.hk}

\author{Zhicong Lu}
\orcid{0002-7761-6351}

\affiliation{%
  \institution{City University of Hong Kong}
  \streetaddress{83 Tat Chee Ave}
  \city{Hong Kong}
  \country{China}
}
\authornote{Corresponding Author}
\email{zhicong.lu@cityu.edu.hk}

\renewcommand{\shortauthors}{Ryan Yen, et al.}

\newcommand{\sys}[0]{{{\it StoryChat}}}

\begin{abstract}
 Live streaming platforms and existing viewer participation tools enable users to interact and engage with an online community, but the anonymity and scale of chat usually result in the spread of negative comments. However, only a few existing moderation tools investigate the influence of proactive moderation on viewers' engagement and prosocial behavior. To address this, we developed StoryChat, a narrative-based viewer participation tool that utilizes a dynamic graphical plot to reflect chatroom negativity. We crafted the narrative through a viewer-centered (N=65) iterative design process and evaluated the tool with 48 experienced viewers in a deployment study. We discovered that StoryChat encouraged viewers to contribute prosocial comments, increased viewer engagement, and fostered viewers' sense of community. Viewers reported a closer connection between streamers and other viewers because of the narrative design, suggesting that narrative-based viewer engagement tools have the potential to encourage community engagement and prosocial behaviors.
\end{abstract}

\ccsdesc[500]{Human-centered computing~Collaborative and social computing systems and tools}
\ccsdesc[300]{Human-centered computing~Collaborative and social computing design and evaluation methods}

\keywords{live streaming, narrative, user engagement, moderation, community, intervention}

\begin{teaserfigure}
    \includegraphics[width=\textwidth]{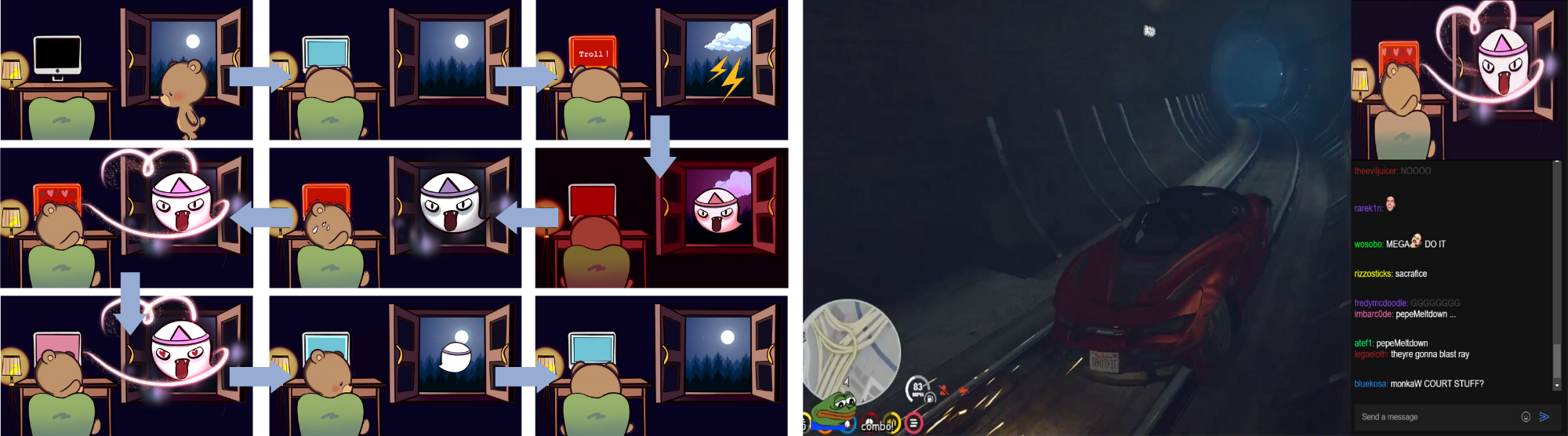}
    \caption{\sys{}, a live streaming viewer participation tool, obtains real-time comments in a chatroom and displays a narrative graphical design representing negative and positive comments. Left: The storyline and graphic design of \sys{}'s narrative design. Right: A screen capture of \sys{}'s interface including the live streaming video (left), narrative design (top-right) and chatroom (bottom-right).}
    \label{fig:teaser}
    \Description{A 3x3 comic representation of the narrative progression for StoryChat, beginning from the baseline happy state, changing to a darker atmosphere and the eventual appearance of the ghost as negative comments start being detected in the live stream chat, followed by a gradual return to the base-state as viewers start sending in positive comments in the live-stream chatroom. (on the left) An image of what the prototype looks like. (on the left) the Twitch live stream. (on the right) the live-stream chatroom, with Storychat embedded on the top.}
\end{teaserfigure}

\maketitle

\section{Introduction}

Live streaming becomes an increasingly ubiquitous form of entertainment over these years, and there has been growing interest in exploring various ways for viewers to engage in the communities \cite{lim2020role, hilvert2018social, li2020systematic, lin2021happiness, de2020live}. Researchers have also developed a variety of viewer participation tools for live streaming platforms in order to improve interaction and engagement within the community via various new communication channels \cite{seering2017audience, lessel2017expanding}. However, a concerning side effect of these tools is the rising occurrence of negative behaviors in the community \cite{seering2017audience, striner2021mapping, lu2018streamwiki}, which means that harmful, humiliating, profane, and toxic comments are spread through the live streaming chatroom, indicating the need for moderation of these viewer participation tools as negative comments are likely to reduce viewers' engagement \cite{collier2012conflict}. Although many live streaming platforms have incorporated systemic moderation methods with human moderators, the growing scale of live streaming communities makes it increasingly difficult for platforms to regulate large quantities of comments efficiently \cite{jhaver2019human, madrigal2018inside, ging2019alphas}. Besides, human moderators usually encounter an intensive workload and are emotionally stressed when moderating public discussion platforms \cite{gillespie2018custodians, wohn2019volunteer, seering2019moderator, gerrard2020behind, steiger2021psychological, dosono2019moderation}. These challenges have increased the interest in the use of automatic moderation, such as chatbots and machine learning \cite{schluger2022proactive}.

While these approaches successfully manage the online community, negative behaviors remain to be a critical concern \cite{AnyKey_2016} because these approaches are primarily \textit{reactive}, which only responding to already-posted antisocial information \cite{schluger2022proactive}. Seering et al. \cite{seering2017shaping} indicate that, in contrast to reactive approaches, proactive moderation could have a substantial positive impact on the behavior of other viewers in the live streaming chatrooms. Proactive moderation could also influence the viewer to engage in prosocial behavior by creating norms \cite{fiesler2018reddit, cai2021understanding, cai2022understanding}. Although research on proactive moderation is limited \cite{seering2019designing}, some studies leverage the idea of persuasive design to activate prosocial interaction \cite{brewer2020inclusion, seering2019designing}, which proactively moderate the online community and encourage viewers to behave prosocially. The use of narratives as a main form of persuasive design was found effective in increasing engagement and addressing harassment in online communities \cite{bisker2010personal, dimond2013hollaback, wang2022highlighting, michie2018her}. Narratives have also been widely used as a non-invasive intervention strategy \cite{gerrig1994narrative, fiske1993social, escalas2004imagine} in psychology. Text-based narratives have often been used to improve feedback quality and decrease the incidence of harsh comments, particularly in online communities and news \cite{wu2021better, van2016linguistic, shen2014stories}. In addition to text-based narratives, the power of visual narrative could increase viewers' empathy \cite{keen2006theory, batson1997perspective, brunye2009you}, and encourage prosocial behavior \cite{batson1997perspective, lamm2007neural}. However, there is limited research devoted to using the narrative through visual graphics in online communities for moderation, although the narrative could function the same way in different media \cite{green2000role, green2002mind}. The combination of narrative structure and story elements can also create a powerful method of conveying information that can assist in mental simulation \cite{taylor1989coping}. Inspired by these findings, our study aims to develop a tool, \sys{}, that uses visual narratives as a persuasive design to prosocially and proactively moderate live streaming chatrooms.

\sys{} was designed as a viewer participation tool that leverages visual narrative to proactively moderate live stream chatrooms by the viewer community. We aim to enhance viewers' sense of community and engagement without constraining their participation and encourage them to actively participate in the ongoing discussion by sending prosocial comments in response to negative ones. We first crafted the narrative and visual design through a viewer-centered iterative design process and then integrated the narratives into \sys{}. To test the moderation and engagement effectiveness of \sys{} on live streaming viewers, we deployed the visual narrative in a Twitch-like interface with 48 volunteer viewers to simulate the experience of viewing real-time live streams. We carefully selected the live streams in the video game or just chatting category based on the number of viewers and the frequency of comments determined by the pilot study. We then investigated how viewers perceived this viewer participation tool through a deployment study and semi-structured interviews with viewers that focused on (1) How viewers interacted with the narrative design, (2) If \sys{} effectively increased prosocial behavior in a live streaming chatroom, and (3) How \sys{} affected viewers' engagement and their connection to the community.

Our key findings suggested that \sys{} enhanced viewers' engagement by involving them in changing the storyline, increasing both the degree of entertainment and sense of community during live streams. Additionally, participants reported a greater sense of community and connection as \sys{} encouraged them to prosocially engage with the conversation and play a more proactive role in changing the chat's environment when they observed negative comments. \sys{} makes the following contributions to the CHI and the live streaming community: 

\begin{itemize}
    \item A new interaction paradigm that employs visual narratives to motivate prosocial behavior and increase viewers' considerations toward a live streaming community.
    \item Empirical evidence demonstrating the effect that visual narratives have on enhancing viewer engagement, promoting a sense of community, and increasing social interaction in chatrooms.
    \item Implications on designing a narrative-based viewer participation tool for the live streaming community.
\end{itemize}
\section{Related Work}
\sys{} builds on the findings of previous research on viewer participation tools for live streaming, approaches to moderating online community content, and narrative research that underlies our approach.

\subsection{Viewer Participation in Live Streams}
In recent years, research on viewer participation has shifted the focus from traditional media to live streaming scenarios \cite{hamilton2014streaming, seering2017audience, striner2019spectrum} and from passive spectatorship to active participation \cite{smith2013live, tang2016meerkat}. As a result, several researchers and live streaming platforms, such as Twitch, have suggested various viewer participation techniques to improve interactivity and promote social interaction \cite{cerratto2014understanding, glickman2018design, hamilton2014streaming, seering2017audience, striner2021mapping}.
Some approaches have attempted to improve the viewer-streamer interaction by designing tools to support emerging forms of community gameplay. For instance, Helpstone \cite{lessel2017expanding}, enabled participants to provide contextual information or hints to streamers. Twitch Plays Pokémon used viewer commands from a Twitch chat stream to control Nintendo games \cite{leavitt2014twitchplayedpokemon, ramirez2014twitch} and CrowdChess \cite{lessel2017crowdchess} supported multiple viewers play chess together against an AI. Apart from games being directly designed for viewers, viewer participation can occur in a variety of other ways, such as chatting via text or special emoticons \cite{hamilton2014streaming, tang2016meerkat}, gifting \cite{hilvert2018social}, clicking likes or hearts \cite{friedlander2017streamer}, polling \cite{cheung2011starcraft, lessel2017let}, subscribing to streaming channels \cite{gros2017world}, moderating chatrooms \cite{seering2017shaping}, or playing games with a streamer in video game stream \cite{lessel2017let}. Recognizing the growing attention that has been devoted to viewer participation techniques, researchers have also begun to investigate viewer experiences and interaction \cite{cerratto2014understanding, striner2019spectrum}, as well as the difficulties associated with designing these viewer participation tools \cite{seering2017audience, striner2021mapping}. One of the challenges encountered in designing these viewer participation tools is the negative interpersonal interaction occurred in the community \cite{seering2017audience, striner2021mapping, lu2018streamwiki}. To address this \cite{seering2017audience}, our research aims to design a viewer participation tool that can both increase viewer engagement and promote prosocial behavior or proactively reduce negative comments in a live streaming scenario.

\subsection{Content Moderation in Online Communities}
Content moderation refers to ``the governance mechanisms that structure participation in a community to facilitate cooperation and prevent abuse \cite{grimmelmann2015virtues}'', and is a significant topic in online communities such as live streaming platforms \cite{haimson2017makes, hamilton2014streaming}. Previous research has developed various moderation techniques, such as blocklist tools \cite{geiger2016bot}, rule-based word filters \cite{jhaver2019human, seering2017shaping, jiang2019moderation}, and regular expression filters \cite{jhaver2019human}. Prior research on assessing content filtering tools has focused on user-level interventions \cite{kiesler2012regulating, seering2017shaping, asthana2018few}. Jhaver et al. \cite{jhaver2019does, jhaver2019did} investigated the impact of implementing community norms and provided implications on improving user attitudes about fairness and posting frequency. This research has identified the theoretical underpinnings of the deployment of user-level interventions to enhance community outcomes. On the other hand, there has been less research on community-level moderation \cite{matias2018civilservant, chandrasekharan2019crossmod, chandrasekharan2022quarantined, chandrasekharan2017you}.

Researchers in HCI have investigated the challenges and drawbacks of the above-mentioned content moderation approaches \cite{im2020synthesized, seering2019moderator, wohn2019volunteer}, and have begun to investigate the influence of sanctions on community users \cite{jhaver2018online, jhaver2019does, jhaver2019did, chang2019trajectories}. For example, Seering et al. \cite{seering2017shaping}, discovered that banning any type of behavior had a negative influence on the frequency of that behavior appearing in subsequent comments. They also highlighted a flaw in current reactive moderation approaches, suggesting that they were ineffective at encouraging prosocial behavior. Although various studies have investigated content moderation techniques in online communities, many focused on the role of volunteer moderators \cite{wang2022highlighting} or social actions conducted by other community members or site users in reaction to negative behavior \cite{mahar2018squadbox, seering2019moderator}. Research on proactive interventions is still nascent \cite{seering2019designing}, and only a few studies investigating the influence of moderation on viewer engagement \cite{wang2022highlighting}. Research examining the proactive prevention of negative behavior has found that norms and rules established by a platform or moderator can influence viewers' to act prosocially in the online community \cite{fiesler2018reddit, cai2021understanding, cai2022understanding}. User interfaces with persuasive designs such as CAPTCHAs \cite{seering2019designing} and the GLHF pledge \cite{brewer2020inclusion} successfully activated prosocial interaction, and proactively preventing negative behaviors in online chats. The benefit of such implicit persuasive designs \cite{lewis2011affective, dijksterhuis2001perception, murnane2020designing, nakaoka2021eat2pic, zhang2021exploring, chiu2009playful, purpura2011fit4life, schultz2007constructive} is that they can be adopted subtly and unobtrusively to effectively modify user attitudes and behaviors without diminishing user engagement. Previous research leverages narratives to successfully promote prosocial behavior and address harassment in online communities \cite{dimond2013hollaback, blackwell2017classification}, providing a fundamental basis for us to design narratives as an implicit persuasive design to proactively moderate online community. Our research aims to explore the effectiveness of introducing a viewer participation tool that can promote prosocial behavior while strengthening viewer engagement and one's sense of community.

\subsection{Narratives as a Moderation Approach}
Narratives can act as a cognitive tool for situated understanding, creating opportunities for mental stimulation and transportation by allowing individuals to develop a mental model of narrative through immersion or relating it with their personal experience \cite{busselle2008fictionality, busselle2009measuring}. With narratives, viewers produce internal behavioral episodes that simulate an event and enable them to imagine actual or potential behaviors as if they were the main character of the episode\cite{gerrig1994narrative, fiske1993social, escalas2004imagine}. These opportunities for mental stimulation or transportation often add persuasive power to the underlying messages in the narrative by immersing the individual in the narrative \cite{green2000role, gerrig2004psychological}. Previous research has suggested that reading narratives about others' unpleasant experiences triggers empathy \cite{keen2006theory, batson1997perspective, brunye2009you}, reduces tolerance for negative behaviors \cite{foubert2007creating} and promotes prosocial behaviors \cite{batson1997perspective, lamm2007neural}. In our work, we frame the design process extended from prior research to elicit empathy from viewers by having the narrative design demonstrate how a protagonist experiences negative comments and how viewers can help by sending prosocial comments. We also allow viewers to control the progression of the storyline because the ability to choose paths or actions within the story makes interactive narratives particularly favorable for mental simulation \cite{guzdial2015crowdsourcing}. Research on the impact of introducing interactivity into narratives suggests that they have a stronger effect on viewer attitudes than more traditional forms of narratives, like oral storytelling or episodes, because they evoke more vivid mental imagery, reduce counter-arguing, and increase the identification and character connection between a protagonist and reader \cite{green2014interactive}.

In terms of online communities, prior research has found that sharing and reading others' narratives that illustrate how they respond to harassment assisted harassed people in adjusting their cognition and emotions toward their experiences \cite{dimond2013hollaback}. The extension of this work applied narratives to understanding online harassment \cite{blackwell2017classification, jhaver2018online} and found that participants draw comfort from sharing their experiences in the narrative. Previous research has also used narratives to assist LGBTQ individuals in coming out when recovering from personal crises and traumatic events \cite{dym2019coming}. Michie et al. \cite{michie2018her} designed a storytelling platform for critical reflection on the issue of abortion and found that narratives raise public awareness due to their ability to motivate social action \cite{ganz2001power}. Wu et al. \cite{wu2021better} utilized negative experience narratives to motivate online community members to provide more useful feedback. Although transportation imagery theory \cite{green2000role, green2002mind} claimed that narrative transportation could work on various forms of media, a growing body of research suggests the benefits of using visual narratives over more expository forms as a tool for persuasive communication through tools like suspension of disbelief and intense absorption as a consequence of engagement with the narrative – both of which can often reduce the likelihood of developing counterarguments \cite{brechman2015narrative, murnane2020designing, slater1997persuasion, csikszentmihalyi1990flow} compared the perception of transportation or mental simulation to a state similar to `flow', and heavy immersion into the narrative allows individuals to reduce cognitive resistance and negative psychological reactance while also allowing them to construct a mental model of the narrative that impacts their attitude and behaviour \cite{moyer2008toward, murnane2020designing}. Although the current narrative moderation research has shown the effectiveness of text-based narratives, there has been limited research employing graphical narratives to promote prosocial behavior in online communities. The most relevant research, by Murnane et al. \cite{murnane2020designing}, created a graphic narrative design to motivate users to engage in physical activities by visualizing their activity progress. In a similar manner our research focuses on arousing viewers' empathy and engage them in the chat prosocially to achieve engagement enhancement through the use of a graphic narrative. 
\section{Iterative Design and System Implementation}
\begin{figure*}[htb]
    \centering
    \includegraphics[width=.8\textwidth]{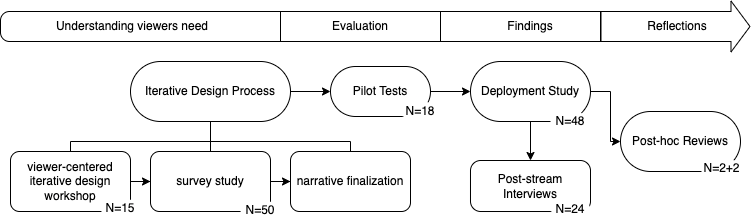}
    \caption{The flowchart of the iterative design process and studies for \sys{}, from left to right, including the low-fidelity design workshop, the survey study on the high-fidelity prototype, the finalization of the dynamic visual narrative, the pilot testing, the deployment study with the post-stream interview, and the post-hoc evaluations from moderators and streamers.}
    \label{fig:flowchart}
    \Description{A flowchart illustrate four phases of design process for \sys{}, including understanding viewers need, evaluation, findings, and reflections. Under these four phases, several studies and design methods being used throughout the study, from left to right, including the low-fidelity design workshop, the survey study on the high-fidelity prototype, the finalization of the dynamic visual narrative, the pilot testing, the deployment study with the post-stream interview, and the post-hoc evaluations with moderators and streamers.}
\end{figure*}

To examine the impact of narratives on viewer engagement during live streaming, we designed \sys{}, which was composed of a narrative graphical design and a Twitch-like interface (Figure\ref{fig:admin-interface}). The system included Twitch streaming content and a real-time chatroom (via Twitch's API) to enable viewers to watch and engage with the chat as they would on an actual live streaming platform. The narrative design was crafted via a viewer-centered iterative design process (Figure\ref{fig:flowchart}), which included a low-fidelity prototype workshop (N = 15), a high-fidelity survey (N = 50), and the finalization of the dynamic graphical narrative. Here, we outline the primary features of \sys{}, including how the chatroom comments affected the narrative animations and the design process used to construct the narratives.

\subsection{Crafting the Narrative: Iterative Design}
\begin{figure}
    \centering
    \includegraphics[width=\linewidth]{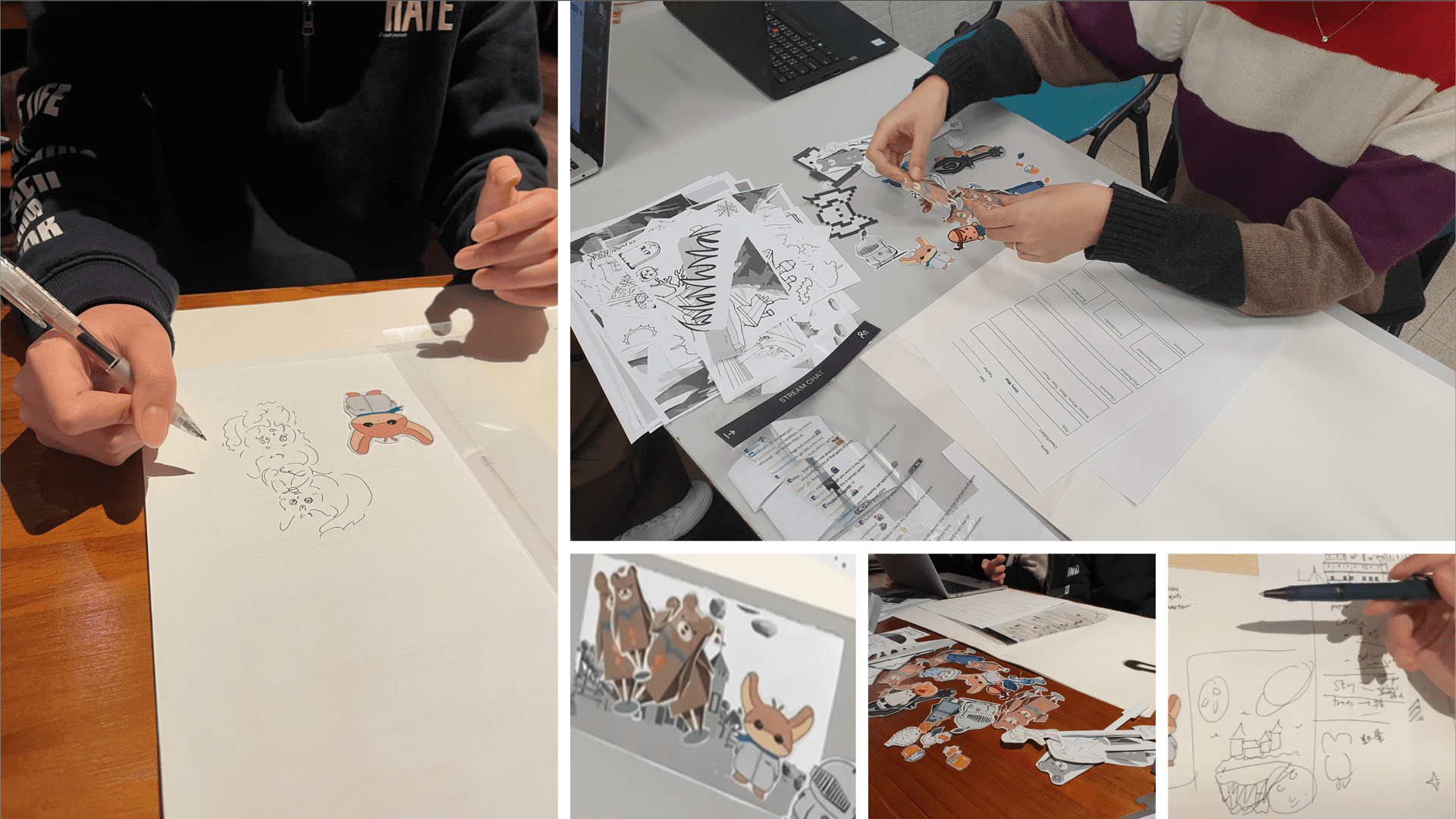}
    \caption{Participants ideation process during the narrative design workshop with experienced live streaming viewers and designers}
    \Description{Five images depicting the narrative design workshop. Participants were encouraged to use pens, paper, and any other tools to develop a narrative to represent the various state changes that could occur in the storyline we designed for StoryChat.}
    \label{fig:formative}
  \end{figure}

To further explore user needs and suitable design factors for the narrative design, we conducted a viewer-centered iterative design workshop with 15 participants (7 female, mean age 22 years, SD = 1.4 years) through convenience sampling. Participants were compensated in accordance with local standards. We posted recruitment messages on social media and sent personal messages to participants that were part of an internal recruitment mailing list. All participants were familiar with live streams and were intermittent to frequent viewers of live streams. Participants were asked to create stories that they believed would be engaging during live streams by sketching or utilizing paper-based elements, which consisted of characters in various emotional states, different locations, background elements, and objects that we provided (Figure \ref{fig:formative}). Participants were asked to design a story whose main plot included the reactions of the protagonist when encountering negative and positive comments. Following this, we conducted semi-structured interviews to understand participants' thoughts on their narrative design as well as their design considerations. From the interviews and observations of participants' design progress, three key design considerations for narratives emerged.

\begin{enumerate}
\item{\textbf{Coherence and Simplicity}}
The majority of participants (N = 11) suggested that they wanted a simple, concise design. This implies that the narrative should be easily understood without the need for explicit explanation, making it universally comprehensible while also ensuring the coherence of the story before and after the occurrence of positive and negative comments.

\item{\textbf{Emotional Resonance: Empathizing with the Characters}} Several participants believed that the style of the narrative would affect their commenting behavior. Some participants claimed that an adorable character could develop emotional resonance (\textit{P4: ``I think cute characters could induce more empathy''}). Participants also emphasized the importance of a logical and understandable narrative that fostered cognitive empathy, stating that the design should immerse viewers in the story so that they can effortlessly imagine themselves in a character's position or circumstances (\textit{P11: ``I think the story should involve the live stream viewers to make it more immersive''}).

\item{\textbf{Symbolic Meanings Within the Narrative.}} All participants carefully considered the underlying meaning and reason behind the use of certain components in their story, often describing the messages behind their design choices (i.e., different emotional states in the character symbolized the mood of the viewers). Participants also suggested the incorporation of a character as a signpost for negative comments in the chat. 
\end{enumerate}

\begin{figure}
    \centering
    \includegraphics[width=\linewidth]{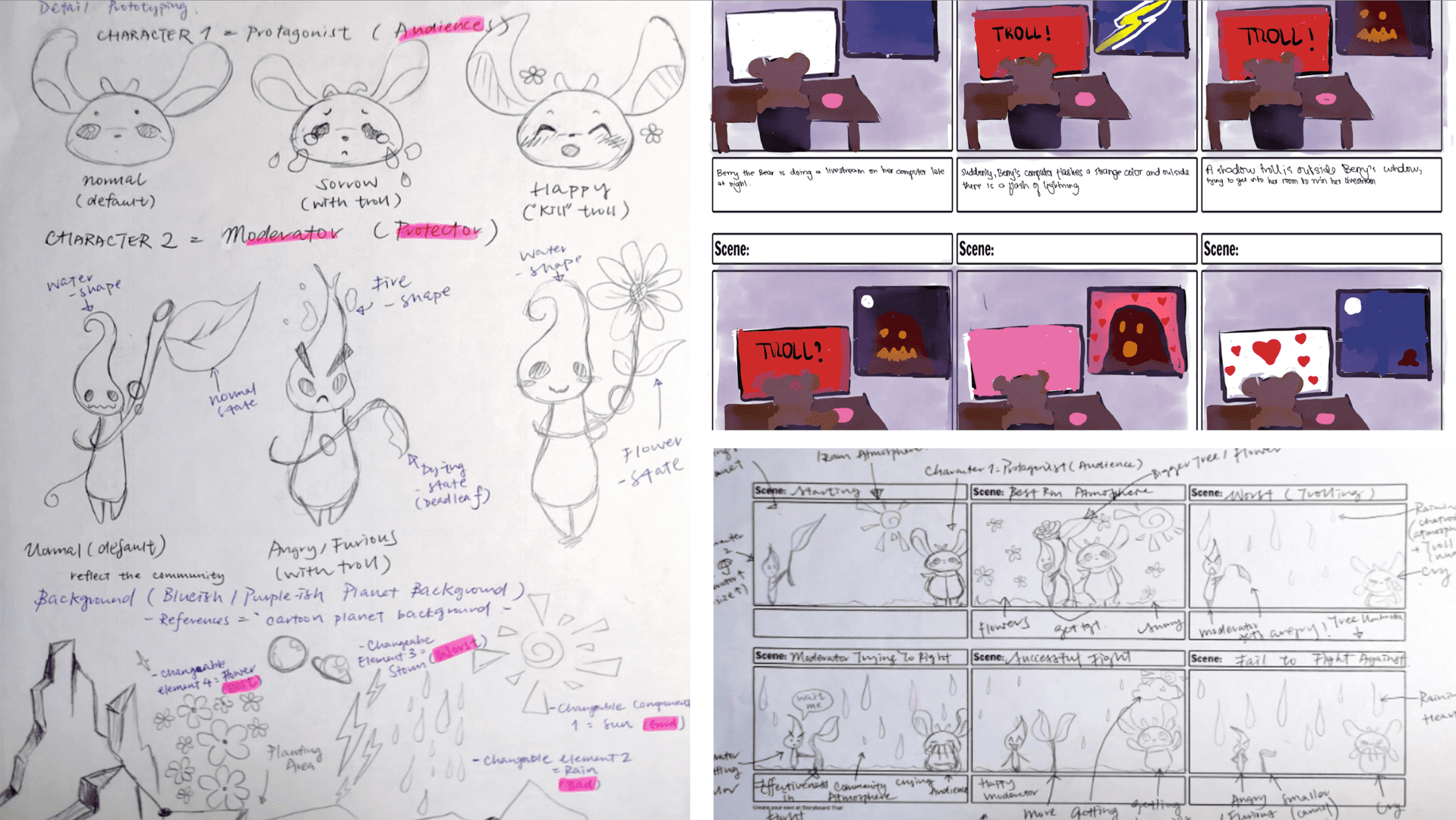}
    \caption{Low-fidelity prototypes of two storyboards representing different storyline development. (bottom-right) Storyboard of \emph{Amazing Plant}; (top-right) storyboard of \emph{Berry and the Ghost}; and (left) detailed settings of \emph{Amazing Plant}. }
    \Description{(on the left) low-fidelity prototype of the storyline ‘Amazing Planet’, depicting the various characters involved in the story. (on top-right) 2x3 storyline for ‘Berry and the Ghost’, involving a series of emotional state-changes meant to reflect incidents in the chatroom. (on the bottom-right) 3x3 storyline for Amazing Planet, which used the concept of ‘growth’ to represent the various state changes in the chat.}
    \label{fig:storyboard}
  \end{figure}

Besides the three design considerations from the design workshop, we identified two types of storyline development: (1) \emph{Comforting the main character - Amazing Planet} (Figure \ref{fig:storyboard} (bottom right). Eight of the participants (P3,5,6,9,10,11,12,13) designed their storyline such that when negative comments were received, the main character would become distressed, whereas when positive comments were received, the character would be comforted. (2) \emph{Driving away the villain - Berry and the Ghost} (Figure\ref{fig:storyboard} (top-right)). The plot was designed by seven participants (P1,2,4,7,8,14,15) such that whenever negative comments arose, a villainous character appeared to threaten the main character. However, if more positive comments appeared, the villainous character would be punished and driven away. Both of these story developments were the result of the viewer's sympathy for the main character, either through a more progressive or conservative approach, in an effort to save or comfort the protagonist.

To compare these two storyline developments, our research team refined the scratch to a higher fidelity design and then conducted a survey study with N=50 volunteer participants (i.e., 21 males, 1 non-binary, 2 prefer not to disclose, ages 18-34, Mean=21) recruited via email, word of mouth, and snowball sampling, who have more than one year of live stream watching experience. According to previous study, paid participants tend to act differently than volunteer participants of live streams \cite{tang2017crowdcasting}. Thus, we centered on recruiting volunteer participants who were intrinsically motivated to watch live streams \cite{lu2021streamsketch, lu2018streamwiki}. Participants were asked to select the most suitable storyline for the live streaming platform and to provide detailed considerations. The survey included likert-scale questions asking participants to rate the design features like background, characters, color scheme, and storyline from 1 (strongly dislike) to 5 (strongly like). In addition, the survey contained the Narrative Engagement Scale \cite{busselle2009measuring}, which measured participants' engagement in narratives from four dimensions: narrative understanding, attentional focus, emotional engagement, and narrative presence. Participants answered 3 questions from each dimension via a 7-point Likert scale ranging from 1 (strongly disagree) to 7 (strongly agree). At the end of the survey, we asked participants to respond to open-ended questions, such as ``Why or why not do you want the design adopted in the live streaming chat room?'' And we did a thematic analysis \cite{braun2012thematic} of these responses to establish initial codes that were refined and integrated iteratively into overarching themes that assisted us make sense of participant feedback on the two storyline developments. From the results, participants overall preferred design B to be implemented in the live streaming chatroom over design A because: \textbf{1) symbolic representation of negative comments.} With the `negative comments' appearing symbolically in the story, participants reported having more empathy for the main character and a deeper emotional connection to the story. \textbf{2) Engagement and Thrill.} The Narrative Engagement Scale reported that design B (Mean=16.21, SD=2.97) received higher score than design A (Mean=7.92, SD=4.37) in all four dimensions. Participants mentioned that the interaction with the villains was more engaging and thrilling because of the way the storyline progressed and their interest in returning the main character of the story to the stable, happy state it was in before the arrival of the villain \textbf{3) Obvious Symbolic Meaning of Characters.} Participants reported they prefer characters in design B (Mean=3.72, SD=0.23) over design A (Mean=2.97, SD=1.2). They mentioned that the characters in design B are more comprehensible, and the symbolic meaning behind each character (i.e., bear is representing streamer) could easily be identified, resonated with, and be closely related to.

\subsection{Dynamic Graphic Narrative: Final Version} 
\begin{figure}
    \centering
    \includegraphics[width=\linewidth]{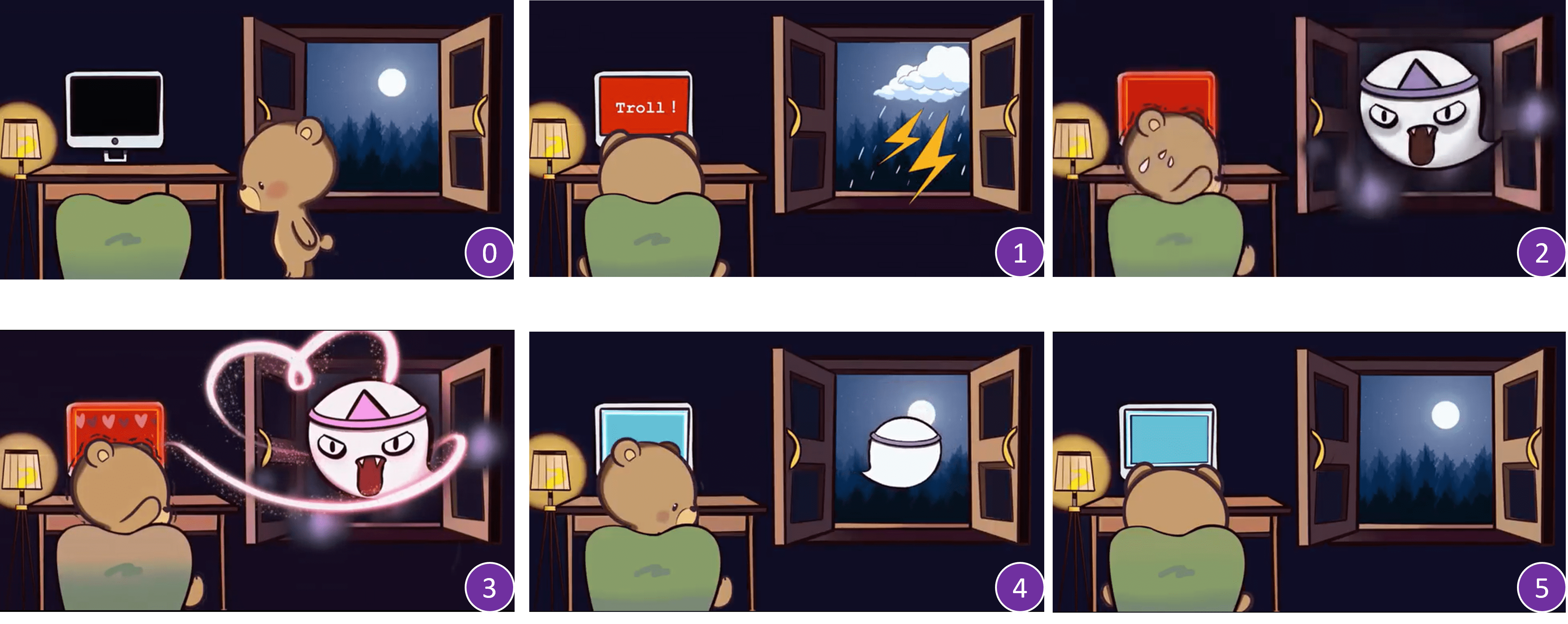}
        \caption{\sys{}'s narrative design can be divided into six major plot points that reflected different events that occurred in the chatroom.}
        \Description{a 2x3 depiction of ‘Berry and the Ghost’, the final storyline, and its progression from 1) stable baseline state to 2) thunder and lightning as negative comments appear in the chat to 3) the appearance of the ghost to 4) the appearance of hearts (meant to represent viewers) battering the ghost to 5) the ghost fleeing and a return to 1) baseline state.}
        \label{fig:narrative-design}
    \end{figure}

In accordance with the findings of the iterative viewer-centered design process, we finalized the dynamic graphical narrative with animations. The Figure\ref{fig:narrative-design} illustrated the major plots in the storyline. The stable state of the story involves a bear character (meant to represent the streamer) interacting with their computer (Figure\ref{fig:narrative-design}. However, as the incidence of negative or toxic comments increase in the actual chatroom, the storyline responds by showing an atmospheric shift - going from a warm, cozy room to depicting thunder and lightning (Figure\ref{fig:narrative-design}. At a certain threshold of negative comments in the live stream chatroom, the symbolic representation of these comments (an evil ghost) appears and the bear reacts in a scared and frightened manner to arouse viewer empathy and encourage them to send positive comments in the chat(Figure\ref{fig:narrative-design}. These positive comments are represented in the form of hearts emerging from the bear's computer and battering the ghost away (Figure\ref{fig:narrative-design}. After a certain number of positive or prosocial comments in the live-stream chatroom, the hearts in \sys{} swarm the ghost and the ghost is chased away - returning the storyline to it's base state (Figure\ref{fig:narrative-design}

This storyline and graphical design aligned with our design considerations: \textbf{1) Coherence and Simpleness.} Some participants in the survey study suggested that the story background distracted their viewing experience. Therefore, we reduced the number of elements in the story as well as ambiguous components in our final design to reduce distraction. \textbf{2) Emotional Resonance: Empathizing with the Characters.} In the survey, 6 participants noted that the original design of the ghost was not attractive and did not match the design style. Thus, we redesigned the ghost and the main character to depict more facial expression. The bear's reaction towards the ghost turned from furious to frightened to be more synchronized with participants feeling and more likely to arouse theirs empathy because of the vulnerability \cite{cohen2001defining}. \textbf{3) Symbolic Meanings behind the Narrative.} Our purpose was to develop an allegory-style story that could correlate participants' behavior to the storyline development. Participants’ reaction towards negative comments further reflect on the protagonist’s action in the narrative. To enhance the participants' emotional resonance, a translucent red mask appears on the screen when the ghost character suddenly appears to signify the danger and evoke visual stimulation for the participants.

\subsection{System Overview and Implementation}
\begin{figure}
    \centering
    \includegraphics[width=\linewidth]{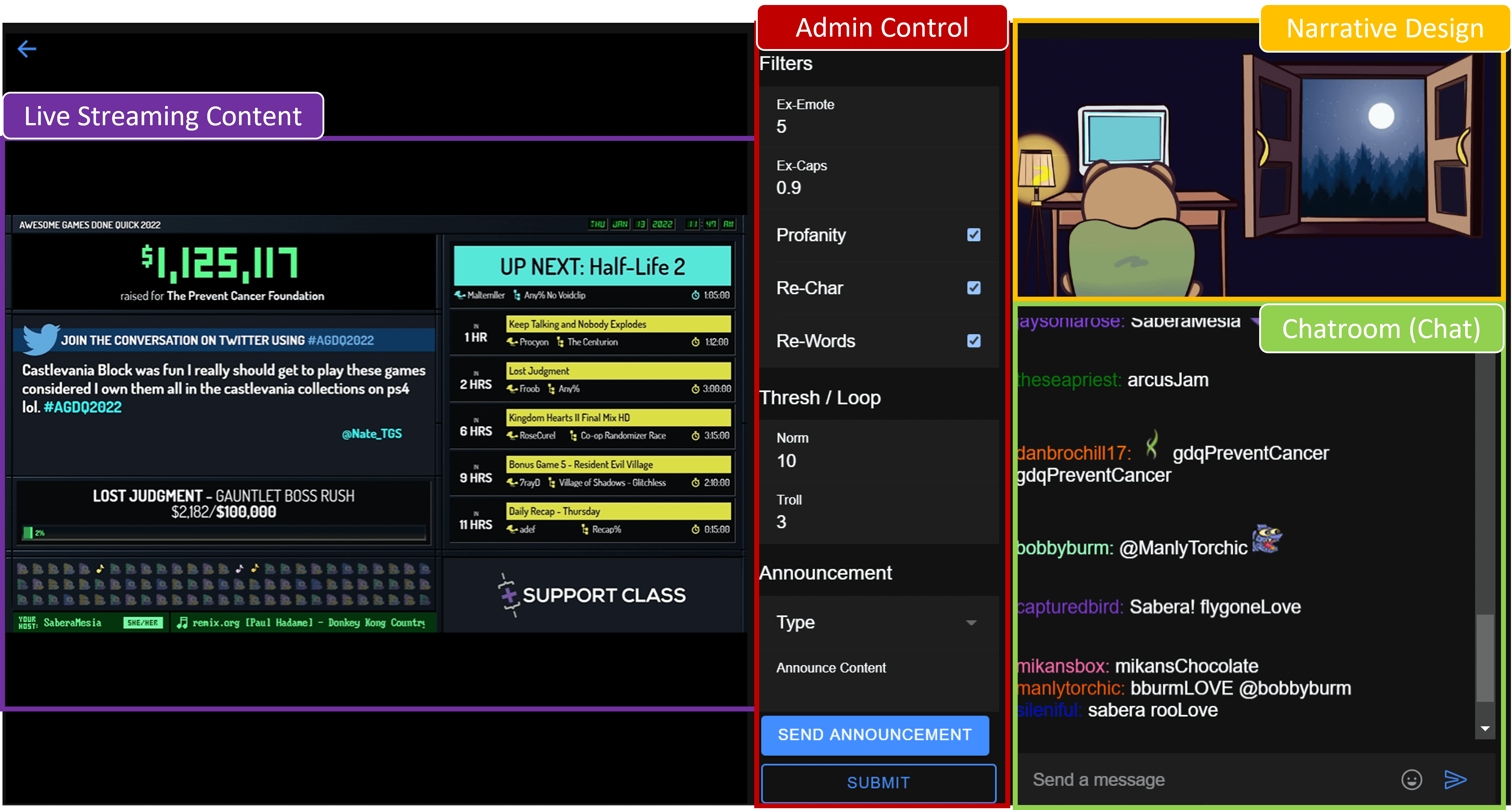}
        \caption{\sys{}'s admin interface with live streaming content on the left, admin control with multiple threshold adjustments in the middle and chatroom with narrative design on the right }
        \label{fig:admin-interface}    
        \Description{A snapshot of StoryChat’s admin interface displaying live-streaming content, threshold adjustments, and a quick view of the chatroom with the embedded Storychat storyline. It contains four sections: live streaming content on the left; admin control with multiple threshold adjustments in the middle; narrative design in the top-right; and the chatroom in the bottom-right side.}
    \end{figure}

\sys{} was developed as a web-based system to allow users to access it on desktop and mobile devices. The system was written in NodeJS under the VueJS front-end framework and used Firebase's database service. \sys{} had four primary visual sections (Figure \ref{fig:admin-interface}): 1) \textbf{Live Streaming Content}, which visualized the concurrent interactive frames of a live stream using the Twitch video API; 2) \textbf{Admin Control}, which is only visible to the admin to change the filter range and negative comment threshold for triggering narrative change. Additionally, there was a broadcasting function that used by the research team to take participants' attendance during the deployment study; 3) \textbf{Narrative Design}, incorporated the visual narrative design via a series of short clips; 4) \textbf{Chatroom (Chat)}, utilized the Twitch Chat API to connect to the real-world Twitch IRC (Internet-Relay-Chat) server. To avoid interfering with real-world chatrooms, we created an additional chatroom that uses the Firebase Realtime Database to connect all participants recruited to our study, while this chatroom will still receive real-time chat from real-world chatrooms via the Twitch API.

\subsection{Negative Comments and Pilot Tests}\label{pilot}
In terms of the mechanisms through which narrative design evolved, we began by defining `negative comments' according to the default setting on one of the most extensively used Twitch conversation moderation bots\footnote{Twitch bot: https://moo.bot/}. By  definition, negative comments were those containing profanity, excessive use of emotes \cite{kobs2020emote, fox2017women}, capital letters, or symbols. \sys{} passed each comment through a comment filter, upon receiving chat from the Twitch IRC server via a socket connection, and determined if it was negative via if-else conditional statement with the definition mentioned above. The classified results affected the narrative's plot change when the number of comments had passed a threshold. To determine the threshold, we conducted six pilot tests with eighteen volunteer participants and collected the number of negative comments in proportion to all other comments. The threshold was set to be 1.12 negative comments per ten thousand viewers every 10 seconds (SD = 1.83), which means that whenever negative comments exceed the threshold in the time window, the narrative plot would change. However, we eventually adapted a more flexible solution by providing the research team a section to adjust the threshold in the admin interface because the distinction between various live streams is obvious. 

Since it is difficult to define `prosocial' comments with existing rule-based or machine learning classification techniques in real time, the narrative plot was controlled via the negativity of the chat room, i.e., when there were less negative comments, the narrative plot shifted to a positive plot (i.e., plot 3, 4, 5 in Figure\ref{fig:narrative-design}). After the study, we conducted a Deductive Thematic Analysis \cite{braun2006using, fereday2006demonstrating} and two researchers classified all comments as either prosocial, negative, or normal.

\section{Deployment Study and Post-Hoc Reviews}
The goals of the study were to understand the effect of moderation on viewers and their sense of community and to understand the degree to which \sys{} improved viewer engagement using a narrative-based tool. To this end, \sys{} was evaluated via a controlled deployment study with six live stream sessions that simulated a real time live streaming experience. We chose to conduct a small-scale deployment study using a `staging instance' \cite{grevet2015piggyback} due to methodological limitations of conducting this study at scale. Previous research has found that larger population samples are needed to thoroughly evaluate the usability of a live streaming tool \cite{lu2018streamwiki, miller2017conversational}, however, a large proportion of Amazon Mechanical Turk-hired viewers tend to be more active than volunteer viewers \cite{tang2017crowdcasting}, which might lead to distorted results. Moreover, as discovered in a pilot study, participants would not actively send negative comments throughout the experiment. Therefore, we created a staging instance with a massive real-time chat copied from actual live streams \cite{chandrasekharan2019crossmod}. Similar to real-world live streams, the study replicated an environment where active viewers could comment and interact with the live streaming community through chat in real-time and experience the mental or behavioral changes that resulted from the occurrence of negative comments. Because the chatroom of \sys{} was linked to an actual Twitch stream, there were viewers of the Twitch streams who were not involved with our study (non-participants) but their comments were presented during the deployment study.

In addition to the deployment study, we conducted a post-hoc review study with two moderators and two streamers to collect their perspectives and reflections on the findings and \sys{}. One of the design considerations for \sys{} is to leverage the concept of `community-led,’ utilizing narrative to influence viewers to make prosocial comments and avoid negative ones. Our research focused on learning how viewers' behaviors changed when using a narrative-based tool, and we did not want the new interface to distract streamers, which would affect viewers' experiences. As a result, we did not design features for streamers or moderators to interact with \sys{}. Although we did not involve streamers and moderators in the deployment study, their reflections about such a narrative-based tool are critical for future design and contribute deeper insights and implications to our findings.

\subsection{Participants} 
Using social media and gaming networks, 48 volunteer participants (24 female, mean age 22 years, SD=2.1 years) were recruited from streamers' Discord channels (N=11) and word-of-mouth and snowball sampling (N=37) within live streaming communities. All the participants had watched live streams for at least six months and watched live streams at least 3 times a week. Participants for this study were limited to the ages of 18 - 24 and recruited largely from university campuses based on Twitch metrics, which suggested that this was the biggest demographic on Twitch \cite{streamscheme_2022}. 

For the post-hoc review study, two streamers (S1 and S2) with more than two years of experience in live gaming streaming and two moderators (M1 and M2) with at least one year of experience moderating live streaming chat rooms and Discord channels were recruited through personal connections. Streamers and moderators were compensated with a \$30 PARKnSHOP Gift Coupon for the approximately 45-minute session.

\subsection{Apparatus, Method, and Procedure} 
\begin{figure*}[htb]
    \centering
    \includegraphics[width=\textwidth]{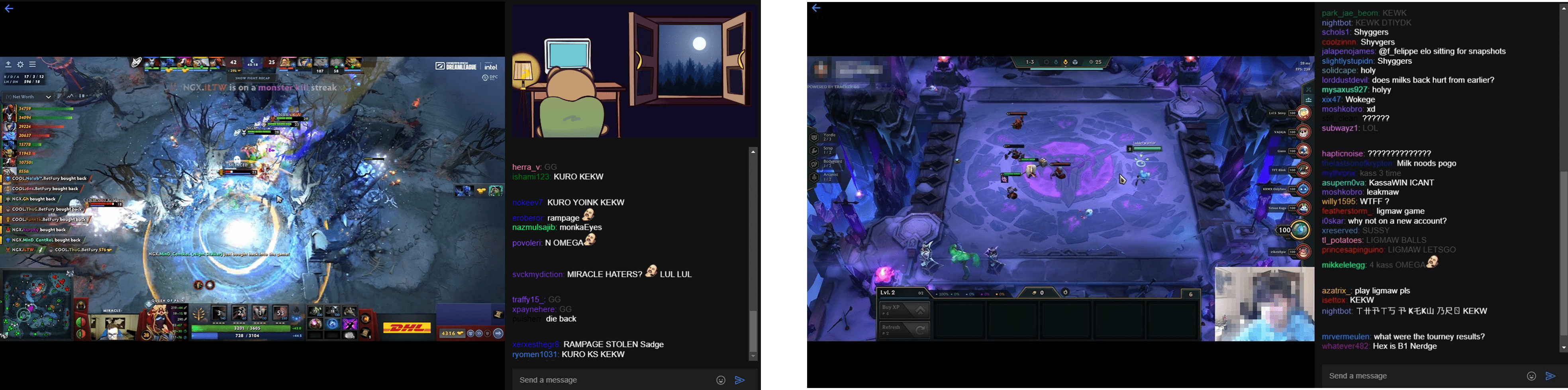}
    \caption{\sys{}'s deployment study which used a with-story mode (Left) and without-story mode (Right)}
    \label{fig:study}
    \Description{A comparative view of the tools used in the deployment study to test the efficacy of Storychat (on the left) the piggyback prototype displaying the Twitch stream along with the embedded Storychat system, (on the right) the normal Twitch interface.}
\end{figure*}

After considering several live streaming platforms, Twitch was selected as the streaming platform for the study because of its community-based structure \cite{seering2020takes}, abundance of large-scale live streams with massive chatrooms, and vulnerability to disruptive behavior \cite{brant2016twitch, seering2017shaping}. To enable participants to experience a change in the narrative design in a short time, we selected six streaming sessions (4 from the game category and 2 from the just chatting category) with an average of 10k viewers and 1.67 comments per second, and we attempted to select chatrooms that contained the similar occurrence rate of negative comments as in the previous pilot test. The 48 participants were evenly distributed among six sessions (i.e., eight participants per session). Four of the participants begin viewing the live stream in the with-story mode (Figure\ref{fig:study} Left) of \sys{}, while the remaining four were assigned to the without-story mode (Figure\ref{fig:study} Right). 

Before the study, the participants received a consent form and a demonstration of \sys{} via a video call that explained how the narrative design changed based on the comments that were made and how the basic interface features worked. Participants were then asked to complete a survey about their live streaming habits and to use their own desktop or laptop computer to conduct the study. Participant participation was implemented and maintained via a web-based system \cite{Ryan_2021} to facilitate the online deployment study. Once every viewer entered the designated streaming session, they started watching live streaming content and interacted with the community through chat for 15 minutes. The research team then asked them to complete a questionnaire and switch to the other mode. Four participants were picked at random from each session to participate in a semi-structured interview, resulting in a total of 24 interviewees.

Following the deployment study and preliminary data analysis, we invited two streamers and two moderators to provide us their perspectives toward \sys{} and reflections on the findings from the deployment study. Before the review, moderators and streamers were given a consent form, information about the study, and they were instructed to freely explore the interface and familiarize themselves with \sys{}'s key features. During the review session, they evaluated deployment study clips, system logs indicating how the study participants engaged with \sys{}, and participants' responses from post-stream interviews.

\subsection{Metrics}
We collected quantitative data from system logs and questionnaires and gathered qualitative data from semi-structured interviews to gain a deeper understanding of the factors that contribute to participants' behavioral changes and sense of community.

\subsubsection{System Log Data}
For each session, we logged every comment submitted by participants and non-participants in the live stream. These comments were saved with a timestamp, the comment type (negative or neutral), and context. All of the logged data was saved in a Firebase Realtime Database in the JSON format. It should be noted that 1) The system automatically categorized the negative comments using the aforementioned classification technique {(} paragraph \ref{pilot}), however the remaining comments were manually coded as neutral or prosocial during analysis. We evaluated several sentiment analysis APIs \cite{hatfield1993emotional, loper2002nltk, manning2014stanford} and machine learning algorithms \cite{reis2020sentiment, barbieri2017towards, kobs2020emote, chouhan2021sentiment}, but the results were either insufficiently accurate or unsuitable for the unique language used in the live streaming chatroom \cite{hope2019hello, olejniczak2015linguistic}; and 2) we collected comments from both participants and non-participants, but only analyzed comments and report on the findings from our own participants.

\subsubsection{Post-Stream Surveys}
We designed two sets of surveys for the with-story mode and without-story mode, both of which used the (1) System Usability Scale (SUS) \cite{bangor2008empirical} to evaluate the usability of \sys{} for live deployment, on a scale from 1 (low) to 5 (high). The (2) Inclusion of the Other in the Self (IOS) \cite{aron1992inclusion} was used to measure the sense of relationship between participants and non-participants during the study. This questionnaire has been adopted in video gaming contexts to comprehend the connection between players or viewers \cite{robinson2022chat, chen2022lagh, culbertson2016crystallize}. Participants were presented pairs of circles with varying degrees of overlap, ranging from just touching to almost completely overlapping. In each pair, one circle was labeled ``I'' and the other ``Other.'' Participants were asked to choose one of these seven pairings to answer ``Which image best depicts your connection with the community?''. The (3) Brief Sense of Community Scale (BSCS) \cite{peterson2008validation} was used to evaluate participants' connection to the community and operationalize sentiments of group membership, need fulfillment, mutual influence, and emotional connection \cite{maruyama2017social}. On a 5-point scale ranging from 1 (strongly disagree) to 5 (strongly agree), participants rated their agreement with eight statements. 

For the with-story mode, participants were also asked to complete the Narrative Engagement Scale, which measured their narrative understanding, cognitive perspective taking, narrative engagement, and emotional engagement \cite{busselle2009measuring} using a 7-point Likert scale ranging from 1 (strongly disagree) to 7 (strongly agree). This scale has been adopted in video and audio narrative scenarios before \cite{richardson2020engagement}. They also completed self-defined 5-point Likert questions which were tailored to the \sys{} and narrative design probed participants' feelings toward \sys{} (Figure\ref{fig:self-likert}).

\subsubsection{Post-Stream Interviews with Participants.}
To gain more insight from participants' connection with community and behavioral changes, we conducted a semi-structured interview (questions in Appendix:\ref{appendix:interview}) to collect participants' perceptions of the differences between using the system with and without story mode in several aspects, including comment sending, sentiment changes, and moderation effectiveness. All 24 interviews were conducted remotely over Zoom, and lasted between 30 to 50 minutes. Interviews were recorded and then transcribed using a paid transcription service\footnote{https://www.iflyrec.com/}. Finally, two co-authors read through the transcripts and coded them using thematic analysis with 201 extracted key phrases.

\subsubsection{Post-Hoc Reviews with Streamers and Moderators.}
The post-hoc review interviews were semi-structured, audio recorded, and think-aloud data \cite{van2003retrospective} was collected from these additional participants. We focused on how they reflected on the results and how they anticipated how the features of a community-led narrative-based tool in a live streaming setting could be enhanced.
\section{Findings}
In this section, we present findings from the deployment study of \sys{} to understand how participants interacted with, engaged with, and perceived the narrative-based participation tool, the extent to which participants changed their behavior, and how the narrative proactively moderated live streaming chatrooms. In addition to the deployment study results, we also report streamers' and moderators' reflections on \sys{} during post hoc reviews. In general, both the deployment study and the post hoc reviews found \sys{} to be engaging and effective in promoting prosocial behavior.

\subsection{Use of \sys{}}
At the beginning of the study, participants were curious and made many different comments and tried to influence the narrative due to the novelty of the system. Participants found that the \sys{} is well integrated and the SUS usability score indicates the \sys{} is easy to use for many participants albeit there is space for improvement.

\subsubsection{System Usability}
The SUS was used to assess the difference between usability of with and without-story mode. We conducted a Wilcoxon Signed-rank test on the result of all SUS questions for the with-story (Mean=70.67, SD=10.46) and without-story (Mean=67.87, SD=13.28) modes. We chose this test because the Likert scale result is interval (normality assumed) ordinal data from two paired groups. Among all the questions, only one showed significance: \textit{``I thought the features in this \sys{} were effectively integrated''} (W =17.0, p < 0.05). This indicates that with narrative design, features in the system were more effectively integrated. Apart from this, all other scales had a p-value greater than 0.05, indicating that the narrative design integrated well with the system interface, without introducing negative impact.

\subsubsection{Novelty Effect}
Participants were first intrigued by this new appearance and sent numerous comments attempting to understand the mechanism behind \sys{}. One user stated that \textit{``I attempted to send a great number of monetary symbols, thinking that it would be a sort of negative comment''} (P3). According to the comment analysis, participants submitted 53.71\% of the total comments, 89.35\% of the negative comments, and 43.1\% of the prosocial comments in the first two minutes. However, considering that this was caused by the novelty effect, we excluded these comments from our quantitative analysis. The participants’ curiosity about how comments impacted the narrative (P1, 8, 12, 14) at the beginning made them contribute more negative comments. Two participants also stated that they would be tempted to send some prosocial or negative comments to see more graphical changes in the narrative design because they felt that it provided a more enriching and fascinating live stream experience. P8 stated that \textit{``if there are no negative comments, I’ll troll to see the story’s response''}, while P14 expressed an interest in seeing \textit{``how bad the story could go''}.

\subsection{Viewers' Engagement and Participation}
\begin{figure}[]{
    \includegraphics[width=\linewidth]{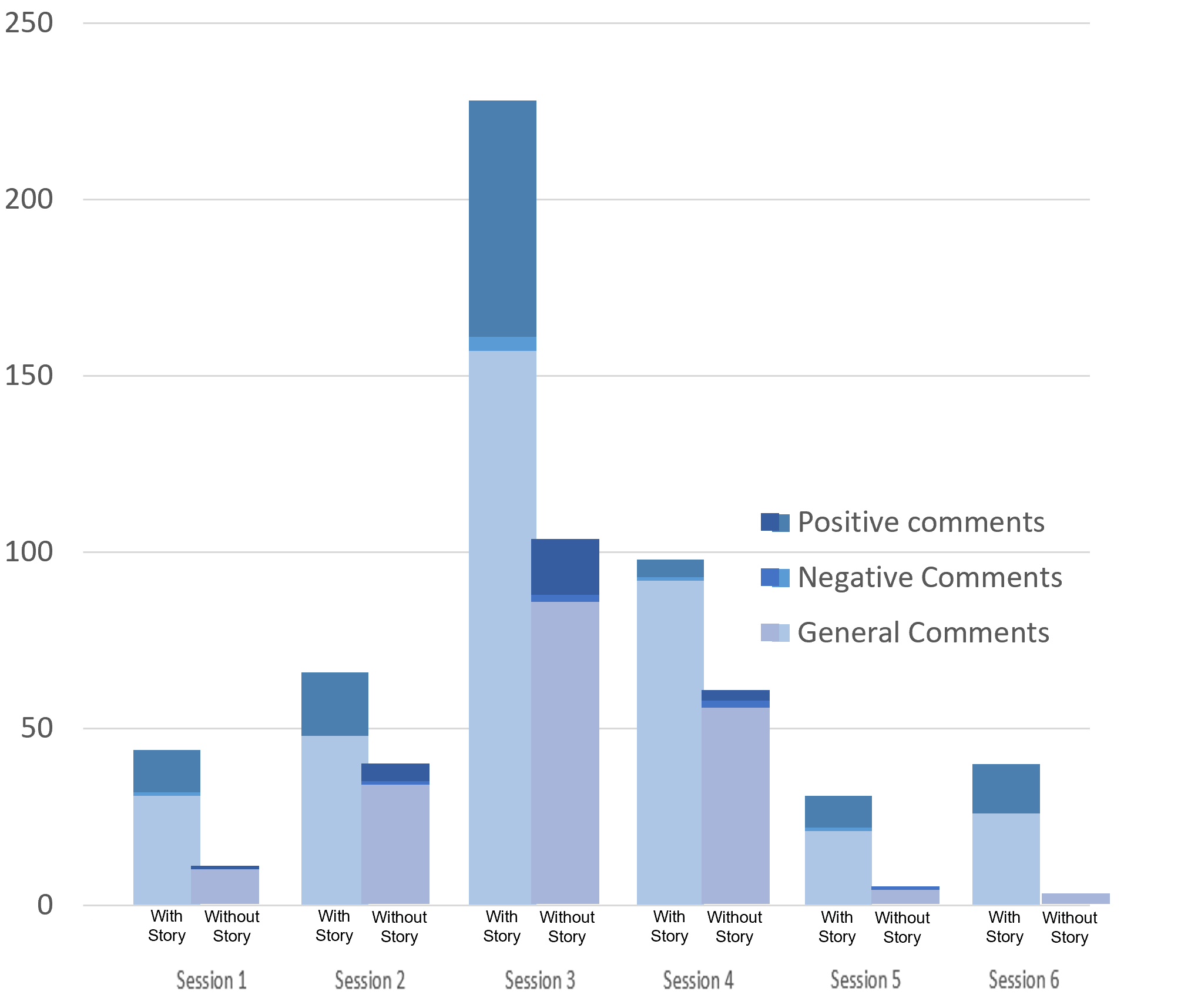}
    \caption{Amount of comments sent by users from the six sessions in two different modes (i.e., with-story and without-story).}
    \Description{A stacked bar graph displaying the number of different types of comments sent by users in either of the two testing modes (with or without Storychat). Through six deployment study sessions, a bar chart depicts the number of comments sent when using with and without-story mode. In every session, the amount of comments send when using with-story mode was higher than when using the without-story mode. Session 1 and 3 had the largest differences, while session 2 had the smallest.}
    \label{fig:avg-comments}
    }
  \end{figure}

When viewing a live stream using the narrative design, participants contributed an average of 84.5 (SD=67.82) total comments, Mean=1.17 (SD=1.34) negative comments, and Mean=20.83 (SD=21.04) prosocial comments to the chatroom (Figure\ref{fig:avg-comments}). In the without-story mode, they provided an average of 37.33 (SD=36.35) total comments, Mean=1 (SD=0.82) negative comments, and Mean=4.17 (SD=5.58) prosocial comments. A paired T-test was conducted on the number of chat comments between the two modes, and the results showed that with the narrative design, participants sent significantly more chat comments (p=0.0286).

\begin{figure}[htb]{
   \includegraphics[width=\linewidth]{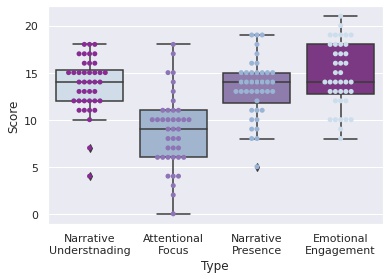}
    \caption{Participants' responses to the Narrative Engagement Scale.}
    \Description{ A box and whisker plot displaying the spread of responses sent in by participants for the Narrative Engagement Scale, showing the means and SD of four scales of the Narrative Engagement Scale, which included narrative understanding, attentional focus, narrative presence, and emotional engagement. Emotional engagement, narrative presence, and narrative understanding received similar scores of around 13 to 15. While the attentional focus received the lowest score of 8 with the largest amount of variance.}
    \label{fig:narrative-engagement}
    }
  \end{figure}

The Self-Defined Likert Scale (questions in Appendix: \ref{appendix:likert}) Results revealed that participants were affected by the narrative design (Mean=4.13, SD=0.97) and were helped in interacting more with chat (Mean=4.02, SD=0.83). Although they believed that their comments had little effect on the narrative (Mean=3.43, SD=0.84), they were inspired to respond after seeing the negative comments (Mean=4.18, SD=0.75) due to their concern about the narrative (Mean=4.13, SD=1.04). The level of narrative engagement was also evaluated via the Narrative Engagement Scale (Figure\ref{fig:narrative-engagement}). Participants showed high scores on Narrative Understanding (Mean=13.73, SD=2.88), and Narrative Presence (Mean=13.38, SD=3.06). The majority of the participants were easily able to comprehend the narrative plot (Mean=4.1, SD=1.43), both in the characters (Mean=4.73, SD=1.34) and the story thread (Mean=4.9, SD=0.99). This was supported by the findings of the interview, where all the participants were able to recall the plot of the story when asked to do so, although some were not able to remember all the minute details. However, some participants noted that they lost interest in the narrative after watching it repeatedly, stating that: \textit{``I was interested in it at first, but I lost my interest later since it is monotonous ''} (P15) and \textit{``I was interested in it at first, but I didn't pay much attention to it later''} (P12). This may suggest that the story used in this study was too easy to comprehend. The following are detailed descriptions of emotional engagement and attentional focus, the highest and lowest scores on the Narrative Engagement Scale, respectively.

\begin{figure*}[h!]
    \centering
    \includegraphics[width=0.8\textwidth]{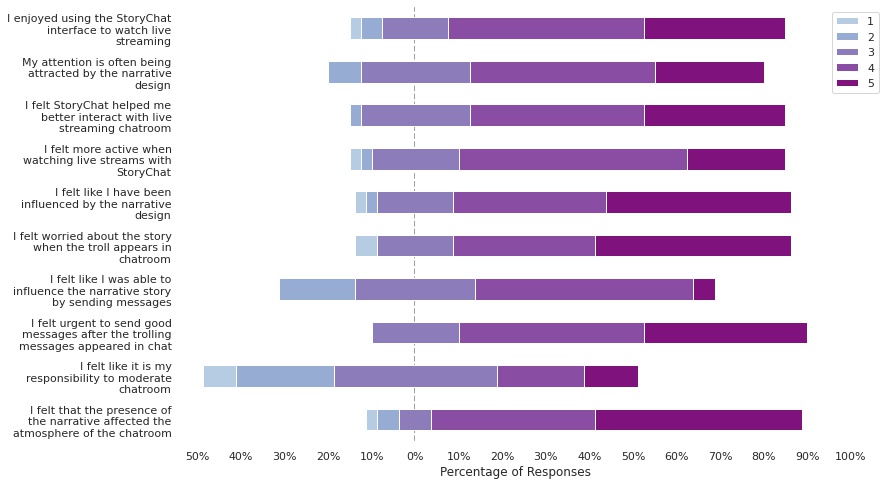}
    \caption{Responses of the Self-Defined Likert Scale questionnaire about the experience of using the narrative design in \sys{}.}
    \label{fig:self-likert}
    \Description{A stacked bar graph displaying the percentage of responses (middle point refers to 3, and 4, 5 would be on the right hand side of the graph with 1, 2 on the left) that agreed with one or more statements about the experience of using Storychat. All the self-defined Likert scale questions are listed in the appendix. This self-defined Likert Scale depicts the responses from users using StoryChat's narrative design. There are ten questions with a rating from 1 (strongly disagree) to 5 (strongly agree). The bar chart shows the percentage of average ratings from 48 participants, where most of the answers were rated  4 or 5, by approximately 85\% of users. However, the question, ``I felt like I was able to influence the narrative story'' only received around 60\% of ratings of 4 and 5. And the question: ``I felt like it is my responsibility to moderate the chatroom'' received the lowest score, where over half of the users disagreed with it. }
\end{figure*}

In general, the narrative design enabled participants to connect with streamers and comment in an engaging manner, and participants believed that the story had an effect on the chatroom's environment (Mean=4.23, SD=0.97) (Figure \ref{fig:self-likert}). They indicated that using the narrative design to watch live streams and interact with the chat was engaging and fun (Mean=4.0, SD=0.96), and they felt more active in the community while doing so (Mean=3.9, SD=0.87). For example, P9 noted \textit{``I felt more engaged with a feeling of participation''} and P11 concluded the reason for the attentional focus was that \textit{``I will take more time to read the chat''}. Another reason might be that they were worried about the narrative development. Many participants stated that they were worried about the main character when negative comments appeared. In addition, we discovered that the majority of participants did not feel they were responsible for moderating the chat (Mean=3.08, SD=1.12), but were instead participating in the community,e.g., \textit{``I believe the story helped me participate more actively in the community, but I do not see myself as a moderator.''} (P12). This result suggests that the narrative design may not lead participants to become moderators or assist in developing a sense of responsibility, but it helped them to participate more in the community and increase their engagement while watching the live stream.

\subsubsection{Attentional Focus}
While the presence of the narrative design made the experience novel and appealing for participants, the majority of participants mentioned that most of their focus remained on the streaming content since that was their primary reason for using Twitch and that they \textit{``focused more on the streamer''} (P6). However, participants acknowledged that watching the stream actually motivated them to check back from time-to-time so they were aware of what was going on in the chat and did not miss anything, paying \textit{``more attention to the chat and story''} than usual (P19), as well as to the \textit{``negative comment and others’ reaction''} (P7) and driving their \textit{``attention to the negative comment''} (P4). Although P5 mentioned that they \textit{``felt a bit distracted since they had to pay more attention to the right-hand side, which is not the streaming content''}, and the Attentional Focus (Mean=8.83, SD=3.87) score related lower in the Narrative Engagement Scale. The findings suggested that the narrative was an attractive, but not disruptive, tool for most of the participants and influenced their community participation but not their live stream watching experience.

Several participants also pointed out that they stopped frequently checking the chatroom and the narrative once they understood the link between comments in the chat and changes in the narrative, only switching their attention back to the chatroom once they observed graphical design changes, which indicated the potential presence of negative comments. In short, participants treated the narrative as a summary or indication of the atmosphere of the chatroom and used it to make sense of the events occurring in the chatroom (P4, 5, 6, 9, 10, 13). P9 stated that \textit{``the most important element for me is the function of summarization''}, while P12 mentioned that they \textit{``don’t usually watch chat, it’s too distracted, but the story helped me to summarize the current situation of the chat''}. This finding was particularly prominent for participants who were not used to viewing or participating in the chat, with many of them pointing out that they were more inclined to pay attention to the chatroom in the conditions with \sys{}.

\subsection{Moderation Effectiveness}
The following section demonstrates how participants moved from becoming emotionally attached to the story, to becoming more connected to the community, to sending more prosocial comments.

\subsubsection{Emotional Engagement and Empathy}
According to the Emotional Engagement questions (Mean=14.73, SD=3.36) from the Narrative Engagement Scale (Figure\ref{fig:narrative-engagement}), participants reported that the narrative affected them emotionally (Mean=4.8, SD=1.3), and they felt the same way the main character (the bear) was experiencing (Mean=5.1, SD=1.28). Some participants could not describe their emotional change in detail, and they just stated that there was some change: \textit{``it affected my emotions at first,''} (P15) and \textit{``There is some change in my mood, but I think it is not nervous''} (P10). Other participants reported several types of emotions, which were summarized into four main categories: curiosity, empathy, irritation, and happiness. Curiosity was one of the main feeling while watching live streams with \sys{} at the beginning of the study. Participants were \textit{``curious about how the story would go when there was a negative comment''} (P16). This curiosity also drove them to influence the story by themselves, e.g., \textit{``I was curious, so I participated in the chat to influence the story''} (P1) and \textit{``I was curious about whether the story plot would change because of my sending comments''} (P12). Some participants were more interested in the principle behind the narrative, expressing curiosity about the \textit{``working principle of the story system ... by the sentiment analysis?''} (P11). Overall, most participants were quite interested in the graphic and storyline changes occurring in the narrative, at least in the first two minutes. 

The interactive narrative also helped participants cultivate a sense of empathy during their viewing experience. Observing the bear protagonist become scared or angry as a result of the ghost antagonist appearing as a representation of negative comments occurring in the chat, led participants to feel sympathetic towards the bear and made them want to engage in actions that could improve the situation for the bear, citing that the \textit{``bear is so poor. It is always scared by the ghost''} (P1) and that they could \textit{``feel that the bear's emotions are not good''} (P5). Increased empathy with the bear protagonist also resulted in increased levels of irritation towards negative comments, especially since these comments were visualized, for many participants. P3 stated that \textit{``this story amplified the annoyance of the negative comment''}, P21 noted that \textit{``I felt annoyed when the ghost comes out, and this feeling was emphasized by the story''} and P6 also suggesting that they were \textit{``annoyed by the same comments repeatedly sent''}. 

Participants also reported feeling happy once the ghost antagonist was expelled from the narrative and the plot returned to its original peaceful setting, with P6 explicitly referencing the impact of the hearts that chased away the ghost antagonist, \textit{``making the audience feel better''} and \textit{``happy that the bad scene was gone''} (P9). P9 also compared their experiences between the two modes, stating that \textit{``after the two tasks, I felt I was happier with the story than without it''}.

An analysis of the chat logs showed a dramatic increase in the number of prosocial comments being sent by viewers right after the ghost antagonist appears in the story (256\% increase ten seconds after the event), suggesting that the main motivation behind viewer behavior was the desire to help the protagonist of the story by eliminating the features that were causing the main character pain or unhappiness. Participants explained their behaviour through statements like \textit{``I wanted to protect the bear''} (P5); \textit{``I wanted to do something for the story''} (P4). Although these motivations seem to be rooted in empathy for the main character of the narrative, a deeper probe into participants' emotional engagement suggested that the ability to influence something greater than the narrative (i.e., the community) inspired participants to continue to engage with the chat and the narrative. This was supported by statements such as \textit{``the story motivated me to type something nice,''} (P20) \textit{``I want to influence the chat,''} (P5) \textit{``I want to participate in the community,''} (P16) and \textit{``I feel that it is my task to improve the chat''} (P9).

\subsubsection{Connecting to the Narrative, Chat, and Community}
While participants' beliefs that their actions were having an impact on the community led many to actively contribute in the chatroom and make a difference in the community, the system did not induce any forced behavioral change, as there were no punishments for not participating in the chat. Participants felt that they were \textit{``not responsible for doing anything''} (P7), but they were more likely to think about the type of comment they wanted to send in the chatroom and the impact it would have on the community. They often choosing to take time to think about what they would like to say, with P12 stating that \textit{``I want to send some comments, but I don't want the story to go badly''}, while P16 said the presence of the \sys{} narrative may \textit{``lower the negative comments, because people will worry when it begins to affect the story''}. This suggests that the presence of \sys{} was a non-intrusive yet reliable behavioral nudge, with participants thinking about, and interacting with, the community more despite not being forced to do so.

\paragraph{Connecting to the Chat}
The results of the survey and the interviews suggest that participants were either influenced by the chat or influenced the chat when viewing live streams in the with-story mode. Participants were able to identify a connection between the narrative design and events in the chat, with P5 summarizing this relationship by stating that \textit{``the chat is reflected in the story and the story could influence the chat. They are connected''}, while P9 succinctly described the relationship as \textit{``the story is affected by the chat''}. Almost all of the participants comprehended the logic behind how the narrative reflected the events of the chat and engaged with it by sending comments and observing changes in the plot. The findings of the qualitative interviews further suggested a marked effect of changes in the graphical design elements of the narrative on participants’ attention towards the chatroom. During plot changes in the narratives, participants became \textit{``curious about what happened in the chat''} (P3) and \textit{``motivated to look at the chat''} (P11). P16 and P19 stated that they were interested in using the narrative to identify the negative comments in the chat, checking the \textit{``negative comment when the ghost came''} or \textit{``who was trolling''}, respectively. These unexpected findings allowed participants to connect more closely with the chatroom and quickly understand the atmosphere and trend of the chatroom discussions in real-time.

\begin{figure}[htb]{
    \includegraphics[width=\linewidth]{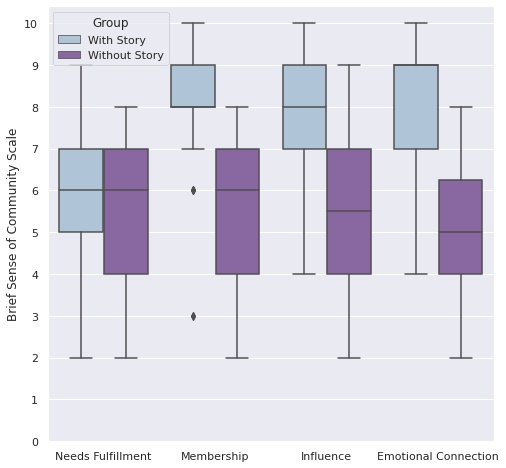}
    \caption{Brief Sense of Community Scale score distribution of with and without-story modes.}
    \Description{A box and whisker plot displaying the score distributions of the Brief Sense of Community Scales across the with and without-Storychat conditions on four different scales: needs fulfillment, membership, influence, and emotional connection. All of the with-story modes received higher scores than the without-story modes, where needs fulfillment had the lowest difference and a similar average score of 6. Among the other three scales, the with-story mode had an average score of 8, and the without-story mode varied from 5 to 6.}
    \label{fig:bcss}
    }
  \end{figure}

\begin{figure}[htb]{
    \includegraphics[width=.7\linewidth]{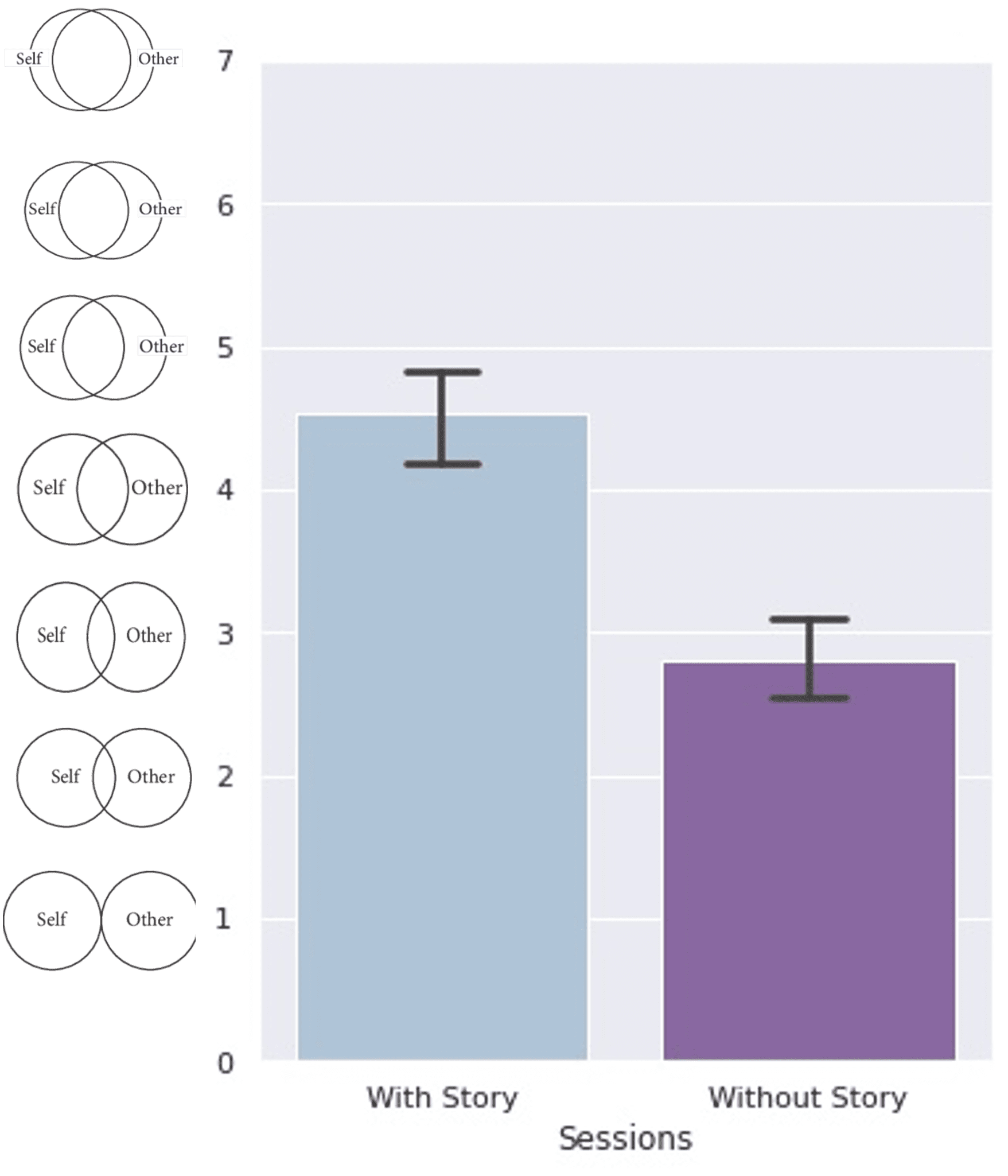}
    \caption{Mean (+/-SE) ratings to the Inclusion of the Other in the Self (IOS) of with- and without-story modes.}
    \Description{ A vertical bar graph displaying the overall ratings for the Inclusion of Other in the Self scale for the with and without-Storychat modes. The higher the score, the stronger the link between self and others, while 0 indicates no intersection. The with-story mode received an average score of 4.5, which outperformed the without-story mode's average score of 2.8.}
    \label{fig:ios}
    }
  \end{figure}

\paragraph{Connecting to Other Viewers}
Findings from the interviews suggest that participants increased level of interaction with the chat by sending comments resulted in feelings of a stronger sense of community and belonging. Although it is difficult to completely quantify such feelings, the IOS scale (Figure \ref{fig:ios}) showed a significant increase from the without-story mode (Mean=2.81, SD=1.31) to the with-story mode (Mean=4.52, SD=1.55). The results from the Brief Sense of Community Scale (BSCS) (Figure \ref{fig:bcss}), reporting higher scores in with-story mode in all four categories, including Needs Fulfillment (with-story: Mean=6.0, SD=1.67 / without-story: Mean=5.68, SD=1.71), Membership (with-story: Mean=8.1, SD=1.61 / without-story: Mean=5.35, SD=1.8), Influence (with-story: Mean=8.0, SD=1.16 / without-story: Mean=5.4, SD=1.79), and Emotional Connection (with-story: Mean=8.18, SD=1.67 / without-story: Mean=5.05, SD=1.86).

The results suggested that participants did feel a sense of inclusion and belonging to the live stream viewer community as a result of the presence of the narrative in \sys{}. P13 contrasted their experiences of watching the live stream with and without-story modes, noting that she was \textit{``upset when no one responded to my comments. The story makes it better and more interactive''}. Participants also enjoyed the sense of camaraderie that occurred as a result of all of the participants working towards the same goal, i.e., eliminating the negative comments, e.g., \textit{``the community can help drive the ghost away''} (P16) and \textit{``I felt that the community and I win the fight together when seeing the ghost expelled''} (P23). Other participants reported feeling more connected to other viewers in the community in the presence of the narrative (P5, P6, P9, P19), with some participants explicitly stating that the presence of the narrative strengthened their interaction with the chat and the community, encouraging them to send more comments to make a difference to the atmosphere in the chat (P1, P6). Although some participants were uncertain about the extent of their impact on the community, believing that their main focus was to alter the narrative (P5, P12), the majority of participants strongly believed that their actions influenced both the community and the narrative. P10 stated that \textit{``I definitely influenced the community since I was sending comments''} and P23 noted that \textit{``positive comments are good for the chat and us''}. Some participants also believed that they were indirectly protecting the streamer by sending prosocial comments into the chat, with P1 believing that their actions could help them \textit{``protect the streamer and the community''} while P13 believed that sending prosocial comments in the chat \textit{``changed the atmosphere because they protected the streamer''}. Participants also reported a sense of achievement after sending a few pleasant or friendly comments to the chat, which inspired them to continue sending prosocial comments and improve the community once they started receiving similar responses.

\subsubsection{From Thought to Action}
A notable impact of narrative was that it not only motivated participants to consider elements of community and togetherness, but it also encouraged them to put them into action. Once participants gained familiarity with the system, the majority of the participants sent prosocial comments as a way to support and protect the community and the streamer against negative comments. Prosocial comments were more prevalent in the with-story mode than in the without-story mode (Figure\ref{fig:avg-comments}), However, it is difficult to discern the difference in the quantity of negative comments since, after the first two seconds, there are merely no negative comments in both with and without-story mode. This finding is consistent with our pilot test observation and past research \cite{glickman2018design} indicating that participants in the `research study' were unlikely to provide negative comments. Although there was no discernible effect in terms of the number of negative comments, participants stated that they were less likely to send negative comments due to the empathy evoked by the narrative and the community's awareness, e.g., \textit{``I aware that people don't swear as much as usual. I think maybe it's because we don't really want to affect others and let the story turn bad because of us.''} Participants noted that the narrative had a positive impact on them as well, with their \textit{``behaviour being affected and influenced in a good way''} (P13), and that this gave them a chance to \textit{``strengthen their communication with others''} (P1). Two participants elaborated on this by stating that although they were not able to change the behavior of the negative comments, they could still improve the live stream viewing experience for other viewers by sending prosocial comments with others and adopting a \textit{``haters gonna hate''} attitude towards the negative comments (P14).

\subsection{Reflections and Perceptions from Streamers and Moderators}
Both streamers (S1, S2) and moderators (M1, M2) found the idea of leveraging a graphical narrative as an implicit and proactive moderation tool interesting and valuable in terms of setting norms in the chat and increasing engagement. However, they also suggested a few limitations of doing so, which needed to be addressed in the future.

\paragraph{Usability, Engagement, and Community Sense}
Both the streamers and the moderators noticed that both prosocial and negative comments increased in the first few minutes. While they understood that this is a common phenomenon when a new extension is applied to a chatroom, the novelty effect is more severe with \sys{} than chatbots. S1 suggested that this might be caused by the visualization involved, i.e., ``\textit{the appearance of the story on top of the chat allows viewers to see the visual change based on what they just sent without the need to consider consequences}''. They also recognized that viewers tended to be more interactive than usual in the with-story mode, which they believed was a positive indication they were engaged in both the community and the chat. However, both S1 and S2 indicated that most of the chat was related to the storyline, suggesting that they might possibly be distracted from the live streaming content, e.g., ``\textit{Although more comments were sent by these viewers, I feared that they were actually more engaged with the chat than with our streaming content}'' (S2). Both streamers and moderators also noted that viewers participated more in the chat and built stronger relationships with other viewers, which is an important factor in feeling more connected to a streaming session and being more willing to participate in the future (e.g., ``\textit{I love the bonding between them (the viewers) as they fight the ghost together}'' (S1) and ``\textit{Although the quality of the stream is the most important thing, belonging to the community that developed through the story is also valuable when speaking to attract viewers to our channel}'' (M1)).

\paragraph{Moderation Effectiveness}
Although the streamers and moderators observed that the disparity between the number of negative comments made by modes with- and without-story did not vary, they were surprised to observe an increase in prosocial comments, which encouraged more viewers to participate in the chat and avoid negative comments, e.g., ``\textit{I believe it is a great achievement that most of the moderation tool I used before can only decrease amount of negative comment and might even decrease the volume of chat after ban}'' (M1). However, both moderators noted that comment removal is still required, since some viewers continued to propagate toxicity in the chat and might irritate other viewers or the streamer. M1 also mentioned that the story facilitates the community's pursuit of justice, which is what most moderators and viewers anticipate when unpleasant comments are made.

Moderators also reported that the narrative would reduce the mental and emotional toll of moderating because moderators often play the role of `bad guy' between the interactions of viewers and streamers. The narrative design could divert viewers' attention, and the protagonist (bear) could serve as a role model in the chat, which is typically the moderators' responsibility. This was supported by statements such as ``\textit{it's nice that I can hide behind the story and communicate with the chat like a normal viewer}'' (M2) and ``\textit{the bear has replaced me as the good example in the chat, which means I don't have to act as the `good student' that everyone hates in school}'' (M1).

M2 mentioned that the narrative would allow him to spend more time on the chatroom or answering questions from viewers, e.g., ``\textit{I would prefer to join in the chat and answer viewers' questions rather than spend too much time banning users or setting up moderation rules}'' while M1 would prefer to maintain their professionalism by keeping a distance from the viewers and focusing on moderation. M1 also raised concerns about the idea of content moderation since the tool only validates the comment and treats all viewers the same, e.g., ``\textit{I believe the design, like most chatbots, has a flaw in that it does not consider that the importance of different comments varies. If the negative comment is from a superfan, for instance, we don't usually ban it as it's most likely to be some sort of joke.}'' This suggests that chat is often tolerated differently by different viewers and that identifying violators is significantly more difficult than distinguishing them based on their comments.
\section{Discussion and Implications}
This study has laid the groundwork for future research on the effects of visual narratives on prosocial behaviors and highlighted several future directions and challenges for improving viewer engagement and viewer sense of community.

\subsection{Engagement and Empathy Arousal}
The results demonstrated that participants were emotionally engaged in the narrative and were motivated to behave prosocially. One possible explanation is that the participants built a mental model of the story plot points by combining their own beliefs, emotions, and goals with the story they are watching \cite{busselle2009measuring, busselle2008fictionality}. Previous studies demonstrated that such narrative involvement might have a substantial effect on the behavior or attitude of viewers \cite{dahlstrom2010role}. The \textit{Extended Elaboration Likelihood Model} offers an alternative explanation for our findings, namely that narratives may drive individuals to engage in prosocial behavior with less cognitive resistance \cite{slater2002entertainment, moyer2008toward, quick2007further} and the more enjoyment with a narrative, the greater the likelihood of overcoming negative psychological barriers \cite{cohen2001defining}. Our findings also revealed that viewers identified with the protagonist and claimed they were driven by a desire to help. This is congruent with the \textit{Entertainment Overcoming Resistance Model}, according to which a `vulnerable' character might evoke a higher emotional response \cite{cohen2001defining, moyer2008toward}. These findings open up new avenues for designing and exploring narratives that can engage and immerse viewers on a deeper level, particularly by incorporating themes related to real-world issues such as racism, sexism, or harassment.

Our results have also shown that proactive moderation provided by the narrative could increase user participation and engagement with the narrative and others in the community. This finding is consistent with previous work, which suggested that positive feedback can encourage users to become more active \cite{schneider2011understanding, wang2022highlighting} and more engaged in the online community \cite{burke2009feed, fanfarelli2015understanding, lampe2005follow}. This engagement could also translate into ``on-the-ground'' activism \cite{lee2013does}, encouraging viewers to post prosocial comments in the chatroom. This finding is similar to previous research on implicit persuasive design \cite{michie2018her}, which suggested that the more users engage with the design, the more likely they are to develop empathy and engage in prosocial behavior \cite{batson1997perspective, lamm2007neural}. Future work can leverage these feedback loops to encourage overall shifts in attitudes in the chatroom, bring about proactive behaviors in online communities, and examining how narratives can affect individual viewers.

In addition to narrative engagement, our findings revealed that empathy was another critical component in promoting prosocial behavior. The results were consistent with previous research, suggesting that narratives enhance the quality of feedback and decrease negative feedback \cite{wu2021better, batson1997perspective, keen2006theory, lamm2007neural}. The ability for narratives to elicit empathy and encourage prosocial conduct is supported by studies on perspective taking, which claim that when viewers are engaged in a narrative and envision how the protagonist feels, empathy is elicited and altruistic motivation is stimulated \cite{batson1997perspective}. Future research could design tools aligned with these narrative models and theories to investigate viewer motivation in depth. Future work could also extend the research to provide a more participatory narrative or system to viewers, which could increase their involvement and enjoyment \cite{green2014interactive, riedl2013interactive}.

\subsection{Connection to the Community}
In addition to the moderation effectiveness brought by the narrative engagement, our findings indicated that an increase in viewers' sense of community prompted them to consider their comments more thoroughly. This result might be explained by studies on bystander intervention, which emphasized the significance of establishing norms to curb negative behaviors \cite{kiesler2012regulating, naab2018flagging}. Empathic sympathy, a psychological method for encouraging bystander intervention, could enable bystanders to perceive the harm caused to victims (bears) by negative comments and behave prosocially \cite{zaki2019war, konrath2011changes}. \sys{}'s narrative design encouraged viewers to post prosocial comments rather than interfering with negative comments. This finding is similar to the study on cyberbullying in social media \cite{taylor2019accountability}, which indicated that empathic concern predicted the likelihood of participants clicking the `like' button on a cyberbullying victim's post but not other intervention behaviors. Future works could examine the bystander intervention regarding to the effectiveness of narrative-based moderation tool. It is also possible to focus on the examination of bystanders and explore the ways in which visual narratives are or are not able to influence their behaviors. Both IOS and BSCS indicated that bystanders were more interconnected, which increased their accountability and framed themselves as a major factor in bystander interventions \cite{difranzo2018upstanding, darley1968bystander}. Research on in-group framing provides an extra rationale for the efficacy of moderation in our research, i.e., perceptions of others as in-group members may increase prosocial behavior \cite{sturmer2006empathy}. The use of in-group framing may elicit empathy from viewers and enhance inter-group connections \cite{stephan1999role}. Future research may employ the idea of in-group framing to design the narrative-based moderation tools and investigate the efficacy of viewer-in-group framing across time to determine how framing evolves.

\subsection{Community-Led Moderation}
Current methods of content moderation on Twitch are both `visible and invisible', with viewers being able to view the moderation occurring in real-time chat but being unaware of or unable to access information about the backstage processes relating to sanctions or suspensions by human moderators \cite{cai2021moderation}. Although Twitch's reliance on human moderators often allows for the development of a relationship between viewers and moderators through the use of live explanations of moderation or creating accountability through the deletion of content, timeouts, or bans, there are still occasions where moderation cannot occur quickly enough or exceeds the moderator's capabilities \cite{thach2022visible}. Additionally, moderators often choose to dismiss actions that would generally result in moderation either because they consider such behavior as being part of the viewer's persona, they feel they have a good grasp on the viewer's intentions, or they just want to distance themselves from the situation \cite{cai2021moderation}. However, the presence of such behavior may still have a negative impact on viewers, who may not agree with or understand the moderator's rationale. \sys{}'s narrative design provides viewers with the opportunity to actively participate in the chat to `drown out' the toxicity and negativity of other comments by engaging in their own type of moderation, with successful efforts on the part of the viewers resulting in the return of the story to its baseline state. The presence of \sys{} itself acts as a motivator to engage in this type of behavior, with viewers being encouraged to work together to create a harmonious environment in the chatroom in response to watching \sys{} state-changes. Future studies can take the next steps in harnessing the power of \sys{} to inspire prosocial and empathetic behaviors in other online social media communities. This includes exploring the development of dynamic visual narratives based on the nature of feedback, discussion, or comments, to building connections and prosocial behavior in interactions with content creators, moderators, or other users.
\section{Limitations and Future Work}

The current \sys{} was tested as a proof-of-concept in a controlled staging instance. To understand the extent to which narrative-based moderation paradigms can impact and influence viewer behavior, future versions of this study can expand to the use of field observations in regards to how viewers respond to the presence of \sys{} and how well it facilitates user-initiated community-led moderation. It is also important to note that the effectiveness of the moderation tool may change over time, as the platform, content, and viewers evolve. This opens up the scope of this line of research, allowing for the exploration of various methods and design tools to develop a generalizable version of narrative-based moderation. Key limitations identified in this study included the impact of narrative based moderation on viewer's cognitive load, opening up avenues to test whether the use of this tool predominantly increased viewer engagement with the community or acted as a distractor as well as questions about the effect of this tool on specific communities \cite{ren2011simulation}, inviting investigation on the extent to which the narrative is useful for minority communities as well as exploration on specific narrative for such communities.

Additionally, the growing occurrences of targeted negative behaviour like hate-raids and harassment towards women (particularly women of color) and LGBTQ+ groups \cite{gray2017blurring, uttarapong2021harassment} may require the need for narrative structures that take into consideration their experiences. Besides viewers, we acknowledge that the findings may not apply to all live streams as the norms vary in different live streaming communities \cite{cai2021understanding}. The narrative structure tested in this study is uniform and repetitive in the case of longer live streams. In the future, research can expand to investigate narrative designs that are customizable for each live stream. Lastly, while initial findings suggest that \sys{} is successful at encouraging viewers to post positive comments and proactively intervening with viewers' intentions to post negative comments, it is difficult to gauge the impact it has on viewers who are purposely being toxic. We hope that future iterations of \sys{} can be implemented in tandem with moderators to reduce the negative impact of viewing toxic chats while also creating a sense of community.

\section{Conclusion}
In this paper, we developed a novel, narrative-based viewer participation tool, \sys{}, that was designed to encourage prosocial behavior and enhance viewer engagement in live streaming chatrooms. We first utilized an iterative, viewer-centered design process to conceive and construct a captivating, yet simple, interactive narrative with a graphical design that altered the narrative's plot based on the current chatroom comments. We then implemented \sys{} in a controlled staging instance that simulated the Twitch viewing experience connecting to a real-world chatroom. The deployment study analyzed the \sys{}'s usability, the effectiveness of moderating, the viewers' engagement, and their sense of community. Qualitative data from interviews and quantitative results from a questionnaire and system logs indicated that \sys{}'s increased the likelihood of viewers engaging with the chat, particularly in the form of empathetic or prosocial comments and also enhanced the viewers' sense of community with the streamer and other viewers in the live streaming, resulting in increased feelings of responsibility and consideration when sending comments. Overall, we believe that these findings support the benefits of using graphical narratives as a moderation tool for viewer-led community moderation and increasing viewer engagement and set the stage for future works that explore the benefits of encouraging behavior change through the use of dynamic, graphical narratives.

\bibliographystyle{ACM-Reference-Format}
\bibliography{sample-base}

\appendix

\section{Post-Stream Interview Questions}
\label{appendix:interview}
\paragraph{Live Stream General Questions}
\begin{enumerate}
    \item For what purposes do you watch live streams? 
    \item How many times a week do you think you watch live streams on either Twitch, Youtube Live, TikTok, or any other platform?
    \item In general, when you interact with the live streaming chat room, what kind of messages do you post?
    \item Why do you post these kinds of messages and what do you hope to gain from interacting with the chat channel?

    \item In one of the streams, there was a short story playing repeatedly. Can you recall what the story’s plot was?
    \item Can you explain what you think the story was put there for?
    \item Which part of the story did you think was the most important? 
    \item What elements or aspects made you think this part of the story was the most important?
    \item In what ways do you think the story impacted your overall live streams experience?
    \item Can you outline some differences you felt while watching the stream normally or with the story?

    \item How does the story influence you? Why? What in the story influences you? (chat behavior, perception to community)
    \item Can you give us an example of how the presence of the story influences your behavior or perspective towards the live streaming chatroom? Why?
    \item Going back to your response on how often you interact with the chat channel – do you think you would interact with chat or community more because of the story? Why would you like to do this?
    \item Do you think you can influence the community? Why do you want to influence it? How can you influence?
    \item In what ways do you think the story influenced your ability to engage with the chat channel?

    \item What elements of the story were particularly interesting for you? Why? 
    \item What elements of the story would you change if given the opportunity? Would you even want to have a story at all? Why?
    \item Which character were you more interested in - the bear or the ghost? Why?

    \item When and why do you send a message? 
    \item When and why do you pay more attention to the story?
\end{enumerate}

\section{Self-defined Likert Questions}
\label{appendix:likert}
\begin{enumerate}
    \item I enjoyed using the StoryChat interface to watch live streaming.
    \item My attention is often being attracted by the narrative design.
    \item I felt StoryChat helped me better interact with live streaming chatroom.
    \item I felt more active when watching live streams with StoryChat.
    \item I felt like I have been influenced by the narrative design.
    \item I felt worried about the story when the troll appears in chatroom.
    \item I felt like I was able to influence the narrative story by sending messages.
    \item I felt urgent to send good messages after the trolling messages appeared in chat.
    \item I felt like it is my responsibility to moderate chatroom.
    \item I felt that the presence of the narrative affected the atmosphere of the chatroom.
\end{enumerate}

\end{document}